\documentclass[aps,prb,twocolumn,superscriptaddress]{revtex4-2}
\usepackage[english]{babel}
\usepackage{bm}

\usepackage{amsmath}
\usepackage{amssymb}
\usepackage{amsfonts}

\usepackage[dvips]{graphicx}

\usepackage[dvipsnames]{xcolor}
\usepackage[colorlinks=true, citecolor=Cerulean,linkcolor=blue, urlcolor = Cerulean]{hyperref}
\usepackage{diagbox}

\begin{document}
\title{Plasmons in two-dimensional electron systems with infinite and semi-infinite metal gratings}

\author{A.~V.~Nikonov}
\affiliation{Kotelnikov Institute of Radio-engineering and Electronics of the RAS, Mokhovaya 11-7, Moscow 125009, Russia}
\affiliation{HSE University, Myasnitskaya str. 20, Moscow, 101000, Russia}

\author{A.~A.~Zabolotnykh}
\affiliation{Kotelnikov Institute of Radio-engineering and Electronics of the RAS, Mokhovaya 11-7, Moscow 125009, Russia}
\affiliation{HSE University, Myasnitskaya str. 20, Moscow, 101000, Russia}

\author{V.~A.~Volkov}
\affiliation{Kotelnikov Institute of Radio-engineering and Electronics of the RAS, Mokhovaya 11-7, Moscow 125009, Russia}

\begin{abstract}
We analytically investigate the plasmons propagating in a homogeneous two-dimensional (2D) electron system, along the metal grating in the form of a periodic array of strip-shaped electrodes (gates) in the vicinity of the 2D system. We show that in the case of semi-infinite grating, the Tamm plasmon modes localized near and propagating along the grating edge can exist in the gaps of the plasmonic band structure provided that the last gate differs in width from the other gates. When the last gate is wider, the spectrum of the fundamental Tamm plasmon mode lies in the lowest frequency gap, below the first plasmonic band. Importantly, we find the mode to exist in this case only for finite values of the wave vector along the grating. Otherwise, if the width of the last gate is smaller, the Tamm plasmon modes occupy higher frequency gaps, and delocalized plasmons from the first frequency band become the lowest frequency excitations. 
\end{abstract}
\maketitle
\section{Introduction}

It is well known that properties of two-dimensional (2D) plasma oscillations (or plasmons) are greatly dependent on the dielectric features of the medium surrounding the 2D electron system~(ES). When a 2D system is embedded in dielectric material, the plasmons a governed by the square-root dispersion law~\cite{Stern1967}. However, in the presence of a metal electrode (gate) near the 2DES the 'image' charges induced in the gate strongly screen the Coulomb interaction between 2D electrons. As a result, in the long-wavelength limit, the spectrum of such gated plasmons becomes softer, acquiring acoustic behavior~\cite{chaplik1972} with the frequency proportional to the 2D wave vector. 

Experimentally, 2D plasmons were first observed in a system of electrons on a liquid helium surface~\cite{Grimes1976} and then, a year later, in silicon inversion layers~\cite{Allen1977,Theis1977}. To date, plasmons have been studied in a wide variety of conducting 2D systems, including quantum wells~\cite{Kukushkin2003,Muravev2008,Muravev2011,scalari2012,Dyer2013,Lusakowski2016,muravev2015novel,muravev2020collective,Shuvaev2021,Andreev2021}, graphene~\cite{woessner2015,Iranzo2018,Bandurin2018,Bylinkin2019,Koppens2020,Kaydashev2020,bandurin2022} and different 2D materials~\cite{basov2016_Science,low2017polaritons,wang2020,Oliveira2021,pogna2024}. From a practical point of view, such a keen interest in plasmons, especially in systems with metal gates, is stimulated by the promising prospects of their application in the design of detectors and emitters of radiation in giga- and terahertz ranges~\cite{Tsui1980,Dyakonov1993,knap2002,Dyakonova2006, otsuji2008emission, knap2009,Muravev2012,Dyer2012,Muravjov2010}. 

\begin{figure}[h]
    \centering
    \includegraphics[width = \linewidth]{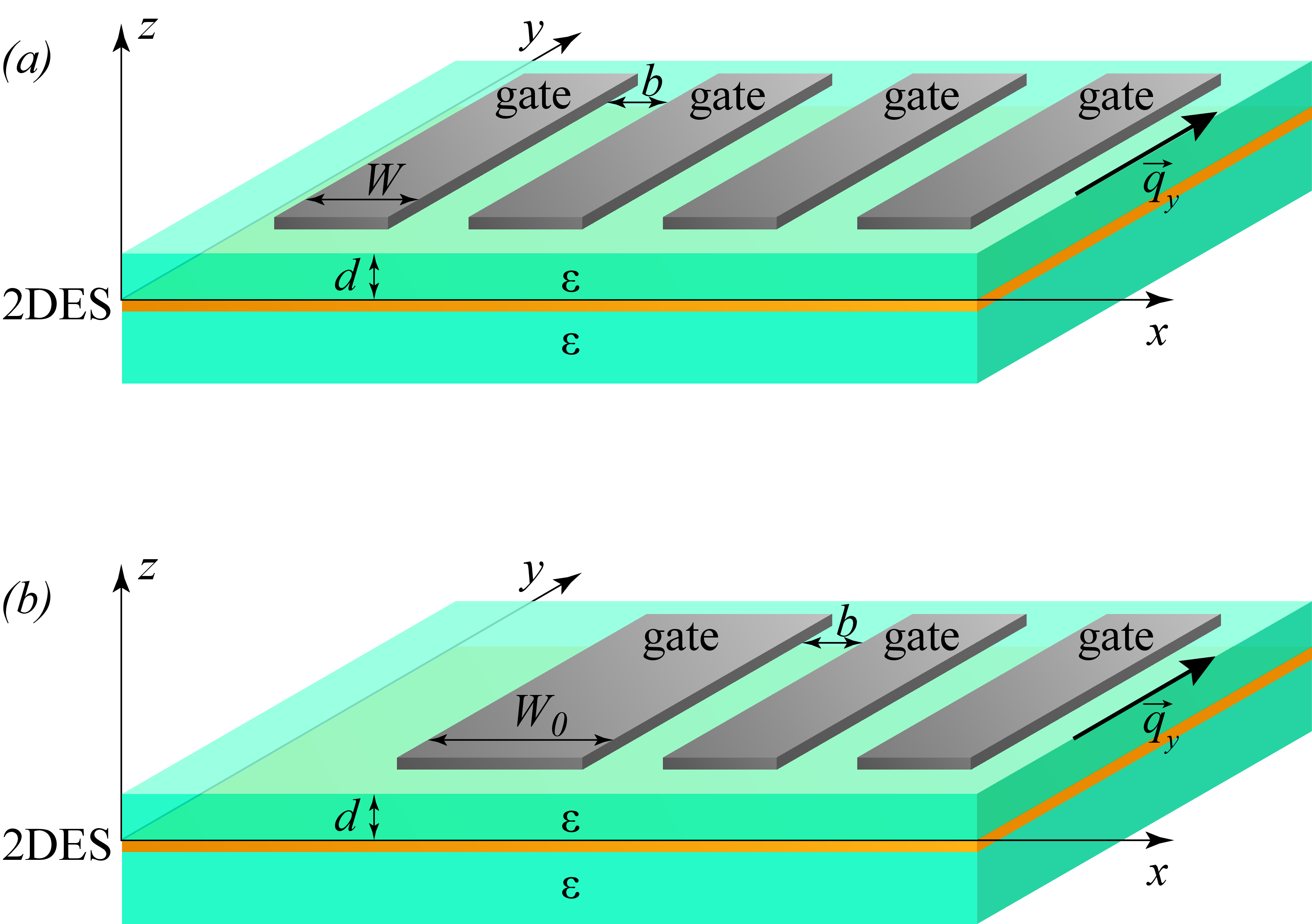}
    \caption{ \label{fig:system} 2D electron systems under consideration, with infinite (a) and semi-infinite (b) planar metal grating gates. The distance between 2DES and the grating gate equals $d$. Media between 2DES and gates as well as below the 2DES have dielectric permittivity equal $\varepsilon$. In Fig. (b) the width of the last gate, $W_0$, may differ from the width of the other gates, $W$. Plasmons under consideration propagate along $y$-axis.}
\end{figure}

The present work is focused on plasma oscillations in a 2DES with an infinite or semi-infinite periodic array (grating) of metal strip gates close to the 2DES, as shown in Fig.~\ref{fig:system}. Plasmons in such systems have been studied in sufficient detail in the literature~\cite{knap2009, Dyer2012, Muravjov2010, Dyer2013,Lusakowski2016, Petrov2017,Bylinkin2019, Kurita2014,Boubanga2014current,Olbrich2016,Boubanga2020,Aizin2023,Sai2023,rappoport2020}. However, to the best of our knowledge, these studies considered only the plasma waves propagating across the strip-shaped gates (in $x$-direction in Fig.~\ref{fig:system}). Therefore, this paper also examines the plasmons propagating {\it along} the strip gates, i.e. with a nonzero wave vector $q_y$, as designated in Fig.~\ref{fig:system}. 

Recently, we have reported on the analytical investigation of plasmons propagating in an infinite and homogeneous 2DES along a single strip-shaped gate~\cite{zabolotnykh2019}, i.e. so-called near-gate plasmons. Unexpectedly, the plasmons were found to be localized near the gate, with their charge density confined almost entirely to the gated area of the 2DES despite the 2DES homogeneity. It was shown that the localization in the direction across the strip-shaped gate is due to the suppression of Coulomb interaction between 2D electrons in the gated area --- the result of image charges being present in the metal gate as opposed to the 'ungated' area of the 2DES. Thus, the weakening of Coulomb interaction causes the localization of plasmons in the gated area of the 2DES. It was established that the fundamental plasmon mode has a peculiar spectrum, combining the features of plasmons in 2D systems with and without the gate~\cite{zabolotnykh2019}. These theoretically discovered plasma modes were observed experimentally in 2D systems based on GaAs/AlGaAs quantum wells~\cite{Muravev2019_2Dstripe,zarezin2020}. Furthermore, they were studied both analytically~\cite{zabolotnykh2019plasmons,zabolotnykh2021} and by experiment~\cite{Muravev_ZarezinPRB2019,muravev2021crossover,zarezin2021review} in 2D systems with a single disk-shaped gate. Note that plasmons propagating along a single gate or a boundary between the gated and ungated regions were also considered elsewhere Refs.~\cite{Dyakonov2008,Petrov2016}.

Thus, the present study is a generalization of the previous works on plasmons propagating in a 2DES along a single gate~\cite{zabolotnykh2019} in a system with metal grating, which is relevant especially to experimental research and possible practical applications. For an infinite grating in the $x$-direction shown in Fig.~\ref{fig:system}(a) we determine a particular band structure for the plasmons and establish its properties. In the case of a semi-infinite grating depicted in Fig.~\ref{fig:system}(b), we analyze the Tamm plasmons that are localized in the $x$-direction near the grating edge and propagate along it. We find that their frequencies lie in the gaps of the plasmonic band structure. We investigate the conditions for the Tamm plasmons to exist as well as their spectral properties and the spatial distribution of the potential.

\section{Analytical approach and principal equations}

Consider an infinite $\delta$-thin 2DES occupying the plane $z=0$, with a grating of planar metal gates at the distance $d$ above the 2DES, as shown in Fig.~\ref{fig:system}. The gates are shaped as strips of infinite length in $y$-direction and finite width in $x$-direction. The infinite and semi-infinite metal gratings under consideration are depicted in Figs.~\ref{fig:system}(a) and (b), respectively. In both cases, all the gates are separated by the same distance $b$ and have equal width $W$ except for the last gate of width $W_0$ in the semi-infinite grating. Throughout the paper, the distance $d$ is assumed to be small enough so that $d \ll b,\,W,\,W_0$, which corresponds to typical experimental conditions~\cite{Iranzo2018,Bandurin2018,Bylinkin2019,Koppens2020, Muravev2012,Dyer2012,Muravjov2010,Muravev2019_2Dstripe, zarezin2020}. The gates are assumed to be perfect conductors with infinite conductivity. In the half-space, $z<d$, the dielectric permittivity is $\varepsilon$, while for $z>d$ it is equal to one. 

To describe plasmons in the system we use the so-called optical approach. It involves finding the plasmon potentials independently in gated and ungated regions of the 2DES and then matching these potentials given the appropriate boundary conditions. This technique is widely exploited to characterize plasmons in various 2DESs~\cite{Satou2003, Petrov2017,Dyakonov2005, Petrov2016,Jiang2018,rejaei2015,siaber2019,Dyer2012, Aizin2012,Dyer2013}, and, as a rule, it provides sufficient description of plasma waves (see also the discussion in Sec.~\ref{sec:DC}).

In particular, we are interested in plasmons propagating along the $y$-axis, therefore we look for solutions in the form $\exp(iq_y y - i \omega t )$. To describe the electron dynamics, we employ a classical method of solving the Poisson equation for the self-consistent plasmon potential, $\varphi(x,z)$, and the 2D charge density, $\rho(x)$:
\begin{align}
    \label{eq:poisson}
    \left( \partial_z\varepsilon(z)\partial_z+ \partial^2_x - q_y^2 \right)\varphi(x,z) = - 4 \pi \rho(x)\delta(z),
\end{align}
along with the continuity equation for the 2D current density, $\bm{j}=(j_x,j_y)$, and Ohm’s law:
\begin{align}
    \label{eq:contin}
    & -i\omega \rho(x)+\partial_x j_x(x)+  iq_y j_y(x)=0,\\ 
    & \quad j_x(x)=-\sigma \partial_x \varphi(x,0), \,\, \, j_y(x)=-\sigma iq_y \varphi(x,0), \label{eq:cur}
\end{align}
where $\sigma$ is the 2DES conductivity. 

Essentially, in the optical approach, the gated and ungated regions of the 2DES are treated as independent and infinite (in $x$-direction in our system) to determine the equations governing the plasmon potential at the 2DES plane, $\varphi(x,z=0)$, in these regions. 

First, consider a gated 2DES. The detailed derivation of the equation for $\varphi(x,z=0)$ for this case is included in Appendix~\ref{app:gated} while the discussion below only stresses the main points. In a gated 2DES, Eq.~(\ref{eq:poisson}) should be solved in $z$-direction, with the boundary conditions of $\varphi(x,z)$ vanishing at the surface of a perfectly conducting metal gate, i.e. at $z=d$, and at $z\to -\infty$. Using Eqs.~(\ref{eq:contin}) and (\ref{eq:cur}) the charge density in Eq.~(\ref{eq:poisson}) can be eliminated to obtain the equation describing only the plasmon potential itself. Furthermore, in the long-wavelength limit, when the distance $d$ is small compared to the characteristic length of potential inhomogeneity in $x$-direction and to $q_y^{-1}$, we arrive at the desired equation:
\begin{equation}
\label{eq:gated}
     \left(\partial_x^2-q_y^2+\frac{\omega^2}{V_p^2}\right)\varphi(x)=0,
\end{equation}
where $\varphi(x)=\varphi(x,z=0)$ is the potential in the 2DES plane and $V_p=\sqrt{4\pi e^2 n d/\varepsilon m}$ is the velocity of plasmons in the gated 2DES~\cite{chaplik1972}, with $-e$, $m$, and $n$ being the electron charge, the effective mass, and 2D concentration, while the dielectric permittivity $\varepsilon$ in the expression for $V_p$ is taken for the medium between 2DES and the gates. It should be noted that in the derivation of Eq.~(\ref{eq:gated}), we apply the dynamic Drude model for the conductivity and assume large electron momentum relaxation time, $\tau$, so that $\omega\tau \gg 1$.

Next, consider an ungated 2DES. The expected frequencies of plasmons are significantly lower than those of square-root plasmons in an ungated 2DES~\cite{Stern1967}, so the latter plasmons are not excited in our system, see also the discussion in Sec.~\ref{sec:DC}. Consequently, it means that the charge density of plasmons, $\rho(x)$, equals zero in the ungated area of the 2DES \footnote{Indeed, plasmons in a 2DES with a single strip-shaped gate were studied in Ref.~\cite{zabolotnykh2019} reporting that the electron charge density is almost entirely localized in the gated area of the 2DES, being absent in the ungated regions, see Figs.~4(a) and (b) in Ref.~\cite{zabolotnykh2019}. For the 2DES with a grating gate considered in this paper, the same behavior of the 2D charge density is expected as the distance $d$ between the 2DES and the gate is small compared to other characteristic lengths, such as $b$, $W$, $q_y^{-1}$, etc. Thus, electrons in the gated areas of the 2DES interact strongly with their 'images' induced in the corresponding metal gates, whereas the interactions of 2D electrons under different gates or electrons in different gates are weak. Therefore, in the limit of small $d$, the description of electron dynamics in the 2DES with a grating gate is very similar to the case of a 2DES with a single gate. It should also be noted that in the framework of the employed (optical) approach, the interaction between the electrons in different gates and 2D electrons under different gates is neglected. }. Thus, at $\rho(x)=0$, Eqs.~(\ref{eq:contin}) and (\ref{eq:cur}) can be reduced to the following simple equation for the potential in the 2DES plane: 
\begin{equation}
\label{eq:ungated}
     \left(\partial_x^2-q_y^2\right)\varphi(x)=0.
\end{equation}

Now, to find the spectrum of plasmons in the 2DES with metal grating, we solve Eqs.~(\ref{eq:gated}) and (\ref{eq:ungated}) according to natural boundary conditions, namely, for the potential $\varphi(x)$ and its derivative, $\partial_x \varphi(x)$, to be continuous at all boundaries of gated/ungated regions of the 2DES. It is worth noting that due to the 2DES homogeneity, the continuity of $\partial_x \varphi(x)$ also leads to the continuity of the $x$-component of the current density across the boundaries of gated and ungated regions. Hence, no charge is accumulated at the boundaries. The given boundary conditions together with the conditions of finite $\varphi(x)$ at $x \to \pm \infty$ are sufficient to determine the dispersion of plasmons in the 2DES with an infinite grating shown in Fig.~\ref{fig:system}(a). To find the spectrum of Tamm plasmons in the 2DES with a semi-infinite grating in Fig.~\ref{fig:system}(b), it is necessary to impose additional conditions of $\varphi(x)$ vanishing at $x \to \pm \infty$, which means that the Tamm plasmons should be localized in $x$-direction in the vicinity of the metal grating edge.

Solving Eqs.~(\ref{eq:gated}) and (\ref{eq:ungated}) with the abovementioned boundary conditions we obtain the dispersion equations for plasmons in the 2DES with infinite and semi-infinite gratings. The explicit derivation of the equations is provided in Appendix~\ref{app:lambdas}. For further analysis, we introduce the 'scaling' factor $\lambda$ that indicates the spatial changes in the amplitude of the plasma wave over one period of the grating $b + W$. The factor can be expressed as follows:
\begin{equation}
    \label{eq:two_lambdas}
    \lambda_{1, 2} = f(q_y, \omega) \pm \sqrt{f^2(q_y, \omega) - 1},
\end{equation}
where
\begin{equation}
    \label{eq:f(q)}
    f(q_y, \omega) = \cos{kW} \cosh{|q_y|b} + \frac{q_y^2 - k^2}{2|q_y| k}\sin{kW} \sinh{|q_y|b},
\end{equation}
with $k^2=\omega^2/V_p^2 -q_y^2$ being an equivalent of the transverse wavenumber for the plasmons.

For an infinite grating the requirement of finite potential at $x \to \pm \infty$ leads to complex-valued factors $\lambda_{1,2}$ in Eq.~(\ref{eq:two_lambdas}), with $|\lambda_{1, 2}| = 1$. This condition is satisfied provided that: 
\begin{equation}
    \label{eq:neq_band}
    -1 \le f(q_y,\omega) \le 1,
\end{equation}
which defines the band structure for the plasmon frequencies.

In the case of the semi-infinite grating gate shown in Fig.~\ref{fig:system}(b), additional modes localized at the edge of the grating are excited, i.e. the Tamm modes. With the introduction of the corresponding factor, $\lambda_{Tamm}$, the dispersion equation for Tamm plasmons appears as in (\ref{eq:disp_tamm_la}). Then, after the substitution of $\lambda$-factor from Eq.~\eqref{eq:two_lambdas} it becomes:
\begin{equation}
    \label{eq:tamm_f(q)}
    f(q_y, \omega) \pm \sqrt{f^2(q_y, \omega) - 1} = e^{-|q_y| b}\frac{\sin{kW_0}}{\sin{k(W_0 - W)}}.
\end{equation}

The localization of Tamm plasmons means that the condition of $|\lambda_{Tamm}| < 1$ is satisfied, which imposes another constraint on the right-hand sides in Eqs.~\eqref{eq:two_lambdas}, (\ref{eq:disp_tamm_la}), and \eqref{eq:tamm_f(q)} as follows:
\begin{equation}	
    \label{eq:con_exist}
    -1 < e^{-|q_y| b}\frac{\sin{kW_0}}{\sin{k(W_0 - W)}} < 1.
\end{equation}
Then, according to Eqs.~\eqref{eq:two_lambdas} and ~\eqref{eq:neq_band}, the condition in~(\ref{eq:con_exist}) implies that the Tamm modes lie in the gaps of the plasmonic band structure. At the same time, it establishes the existence of the Tamm plasmons.

As a matter of clarification, we also make a note of the sign convention used on the left-hand side of Eq.~\eqref{eq:tamm_f(q)}. The sign in Eq.~\eqref{eq:tamm_f(q)} depends on the number of the plasmonic frequency gap occupied by the given Tamm mode. As condition \eqref{eq:neq_band} defines plasmonic bands, so inequality $|f(q_y, \omega)| > 1$ defines plasmonic gaps. Modes that lie in odd frequency gaps have $f(q_y, \omega) > 1$, while for Tamm modes lying in even frequency gaps another inequality, $f(q_y, \omega) < - 1$, is fulfilled. Therefore to satisfy the obtained condition \eqref{eq:con_exist}, one should carefully choose the sign in Eq.~\eqref{eq:tamm_f(q)}, namely: positive or negative sign corresponds, respectively, to an even or odd number of the frequency gap.

\section{Analysis of plasmons in the 2DES with infinite and semi-infinite gratings}

\begin{figure}[h]
    \centering
    \includegraphics[width = \linewidth]{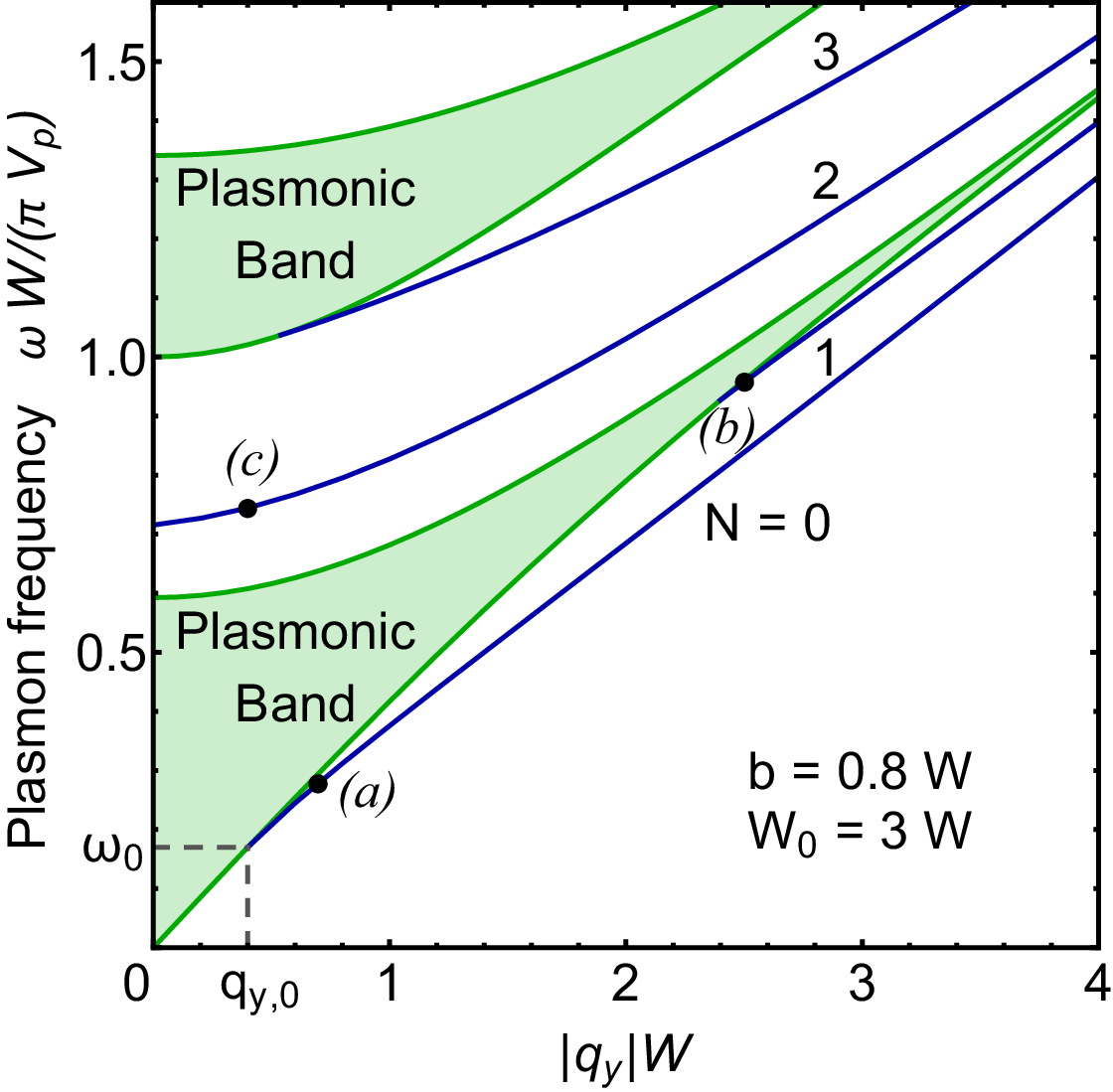}
    \caption{\label{fig:Spec} Spectra of plasmons in the 2DES with a semi-infinite grating gate, as defined by Eqs.~\eqref{eq:neq_band} and \eqref{eq:tamm_f(q)}. The regions shaded in light green represent the plasmon frequency bands. The blue lines denote the Tamm plasmon modes $N = 0, 1, 2,...$. Three black dots $(a)$, $(b)$, and $(c)$ correspond the potentials depicted in Fig.~\ref{fig:Phi}. Parameters $b / W$ and $W_0 / W$ are set to 0.8 and 3, respectively.}
\end{figure}

We begin with the analysis of the plasmonic band structure in the 2D system with infinite grating. In that case, the upper and lower boundaries of the first plasmonic band, $\omega_{up}(q_y)$ and $\omega_{low}(q_y)$, at $|q_y|W \ll 1$ and $b \gtrsim W$ can be represented analytically by the following relations:
\begin{align}
    \label{eq:1band_down_down}
    \frac{\omega^2_{low}(q_y)}{V^2_p} &= \frac{2|q_y|}{W} \frac{ \cosh{|q_y| b} - 1}{\sinh{|q_y|b}}, \\
    \label{eq:1band_down_up}
    \frac{\omega^2_{up}(q_y)}{V^2_p} &= \frac{2|q_y|}{W} \frac{ \cosh{|q_y| b} + 1}{\sinh{|q_y|b}},
\end{align}
where the plasmon velocity $V_p$ is defined after Eq.~(\ref{eq:gated}).

\begin{figure*}
    \includegraphics[width=2.0\columnwidth]{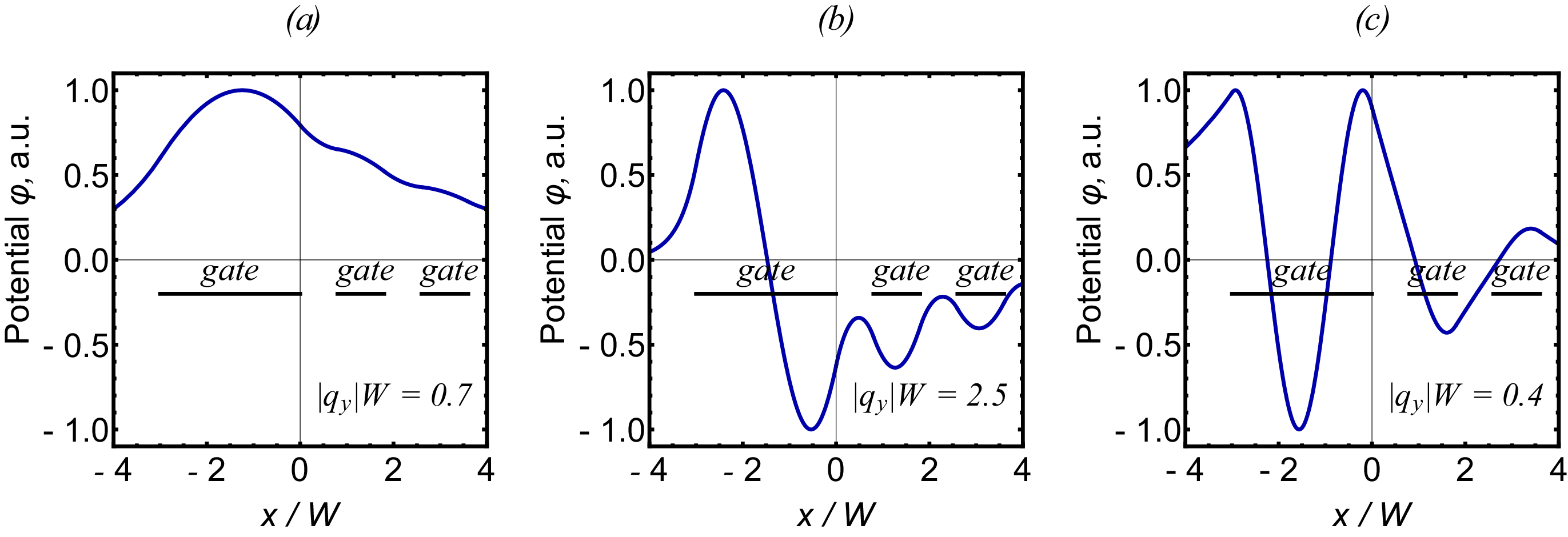}
    \caption{\label{fig:Phi} Potentials in 2DES for Tamm plasmon modes denoted by black points $(a)$, $(b)$, $(c)$ in Fig.~\ref{fig:Spec} are shown. Black solid lines correspond to the location of metal gates. Parameters  $b/W$ and $W_0/W$ have values 0.8 and 3 respectively.}
\end{figure*}

Furthermore, for small separation distances between the gates, when $b \ll W$ (and still $d \ll b$, $|q_y| W \ll 1$), the expressions for the boundaries of the bands of an arbitrary band number $K = 1, 2, ...$ can be derived as follows:
\begin{align}
    \label{eq:Nband_down_small_b}
    \frac{\omega^2_{low, K}(q_y)}{V^2_p} &=q^2_y \left(1 + \frac{b}{W}\right) + \frac{\pi^2 (K - 1)^2}{W^2}, \\
    \label{eq:Nband_up_small_b}
    \frac{\omega^2_{up, K}(q_y)}{V^2_p} &= q^2_y + \frac{\pi^2 K^2}{(b + W)^2}.
\end{align}

Fig.~\ref{fig:Spec} depicts the first and second frequency bands obtained according to~\eqref{eq:neq_band}, where the dark-green lines indicate the boundaries of the bands defined by $f(q_y, \omega) =\pm 1$. In the figure, we can see that in the short-wavelength limit of $|q_y| b \gg 1$, when the wave vector $q_y$ or the distance between the gates $b$ increases, the width of plasmonic bands decreases and becomes infinitely small. At the same time, the boundaries of the band of number $K$ tend asymptotically to the relation for the gated plasmon $\omega=V_p \sqrt{\pi^2 K^2 /W^2 +q_y^2}$. Let us emphasize that the formation of the plasmonic band structure in our study is due to the interaction of currents (and electric fields) flowing in ungated parts of 2DES. These fields and currents are generated by oscillated charge density in gates as well as gated areas of 2DES and can extend into ungated regions over distances up to $q_y^{-1}$, while the charge density in ungated regions is neglected, see Eq.~(\ref{eq:ungated}).

Proceeding to the analysis of Tamm plasmons, we first recognize that if the last gate in a semi-infinite grating has the same width as the other gates, $W_0=W$, the right-hand side of Eq.~\eqref{eq:tamm_f(q)} grows infinitely large, whereas the left-hand side must be less than one. Thus, no Tamm plasmon modes exist under this condition. Conversely, they always appear when $W_0 \neq W$.  The frequency of Tamm plasmons is governed by the wave vector $q_y$ that takes continuous values along the grating as well as by the discrete number $N=0,1,2,..$ corresponding to the number of nodes of the plasmon potential under the last gate of the grating in the transverse direction. Importantly, some of the Tamm modes have a finite frequency at $q_y$ approaching zero. Otherwise, they can reach a plasmonic band at a finite wave vector and frequency. For example, in Fig.~\ref{fig:Spec}, the frequency of the Tamm plasmon mode with $N = 2$ is finite at $q_y=0$, while the modes with $N = 0$, $1$, $3$ touch the plasmon bands at the finite values of the wave vector and frequency. Hence, we find that there are no gapless Tamm plasmons. We also note that at the touching points, the Tamm plasmons become delocalized into the region of the 2DES with the grating $x > 0$.

In the case of $W_0 > W$, the fundamental mode of the Tamm plasmon ($N=0$) always lies in the first frequency gap and terminates in the first plasmonic band at the finite wave vector $q_{y, 0}$ and frequency $\omega_0$ defined by the condition $f(q_{y,0}, \omega_0) = 1$. Then, from Eq.~\eqref{eq:tamm_f(q)}, it follows that:
\begin{equation}
   \label{eq:gap_disp}
   \sin {k_0 W_0} = e^{|q_y|b}\sin{k_0 (W_0 - W)},
\end{equation}
where $k_0=\sqrt{\omega_0^2/V_p^2-q^2_{y,0}}$. As a result, by substituting Eq.~\eqref{eq:Nband_down_small_b} with $K = 1$ into Eq.~\eqref{eq:gap_disp} and applying the conditions $|q_y|W_0 \ll 1$ and $W_0 \gg W$, the expression for $\omega_0$ can be determined analytically. 
At $b \ll W$, we obtain:
\begin{equation}
    \label{eq:width_zero_mode_small_b}
    \frac{\omega_{0}}{V_p}  =\left (1 + \frac{b}{2W}\right) \frac{\pi}{2W_0}\sqrt{\frac{W}{b}}.
\end{equation}

At another limit $b \gtrsim W$ and $|q_y| b \ll 1$ we substitute Eq.~\eqref{eq:1band_down_down} using Taylor series expansion into Eq.~\eqref{eq:gap_disp} and derive the following explicit expression:
\begin{equation}
    \label{eq:width_zero_mode_big_b}
    \frac{\omega_{0}}{V_p} = \sqrt{1 + \frac{b}{W}} 
    \frac{W}{b W_0}.
\end{equation}
Clearly, both asymptotes have a common feature that $\omega_0$ decreases with increasing width of the last gate $W_0$.

Also it is worth investigating a behaviour of Tamm modes with $N > 0$ at $q_y = 0$. In the case of $ b \gg W$ these frequencies lie in plasmonic gaps. One can find the frequencies to be as follows:
\begin{equation}
    \label{eq:zero_qy}
    {\omega_{Tamm}(q_y = 0)} = V_p \frac{\pi N}{W_0}.
\end{equation}

In the long-wavelength limit of $|q_y|W_0 \ll 1$, at $b$ and $W_0 \gg W$ and $q_y b\gtrsim 1$, the dispersion for the fundamental Tamm plasmon mode can be obtained analytically, namely, given $\omega_{Tamm} > \omega_{0}$, we arrive at:
\begin{equation}
    \label{eq:anal_tamm}
    \omega_{Tamm}(q_y) = V_p\sqrt{\frac{|q_y|}{W_0} \frac{2 W_0 - W(1 + \coth{|q_y|b})}{W_0 - W}}.
\end{equation}
It is worth noting that in the limit of the long separation distance between the gates, $q_y b \gg 1$, the spectrum of the Tamm plasmon resembles that of the near-gate plasmon~\cite{zabolotnykh2019} with $\omega^2 = 2V_p^2|q_y|/W_0$. This can be explained by the fact that in the limit of large $b$, the Tamm mode tends to be localized under the last gate, while the contribution of the other gates is negligibly small.

For greater clarity, Fig.~\ref{fig:Phi} includes the partial distribution of potentials for different Tamm modes plotted as a function of the dimensionless coordinate $x / W$. The figure shows that the Tamm modes lying 'deeper' in the frequency gap have shorter localization lengths than those closer to the plasmonic band (compare Figs.~\ref{fig:Phi}(a) and (c)). Another interesting observation is that the potential of the modes from the first frequency gap (with $N = 0$ and $1$ in Fig.~\ref{fig:Spec}) does not vanish at $x>0$, i.e. in the 2DES region with the grating excluding the last gate of width $W_0$, as indicated in Figs.~\ref{fig:Phi}(a) and (b).

As was mentioned above, the fundamental Tamm mode for the case of $W_0 > W$ always exists in the first frequency gap, with its frequency below the boundary of the first plasmonic band. At the same time, this frequency is higher than that of a gated plasmon with the linear spectrum $\omega=V_p q_y$. Therefore, the fundamental mode of Tamm plasmons in this case can be qualitatively considered as an intermediate plasmon mode between those excited in the 2DESs with infinite solid and grating gates.

As for the case of $W_0 < W$, the fundamental Tamm mode lies in a higher plasmonic gap, and the mode may or may not reach a plasmonic band, depending on the relation between the parameters $W_0$, $W$, and $b$. However, our main goal is to investigate the modes with frequencies below the first plasmonic band as these particular modes are more likely to be observable in experiments. For this reason, we do not give detailed consideration to the case of $W_0 < W$, however for a more explicit picture, examples of spectra in this case are given in Appendix~\ref{AppC}.

Nevertheless, we address one peculiar case when $2W_0 = W$, see Fig.~\ref{fig:Spec2}(b). Under such condition, the dispersion equation~\eqref{eq:tamm_f(q)} becomes: 
\begin{equation}
    \label{eq:tamm_f(q)_0.5}
    f(q_y, \omega) \pm \sqrt{f^2(q_y, \omega) - 1} = -e^{-|q_y| b}.
\end{equation}
It is evident that the modes lying in odd frequency gaps (in which the condition $f(q_y, \omega) > 1$ takes place), corresponding to the negative sign before the square root in Eq.~\eqref{eq:tamm_f(q)_0.5}, disappear as the left-hand side of the equation is positive, while the right-hand side of the equation is negative at any wave vector.  On the other hand, the modes in even frequency gaps, corresponding to the positive sign before square root in Eq.~(\ref{eq:tamm_f(q)_0.5}), are excited starting at the upper boundary of the odd plasmonic bands.

Last but not least, it should be noted that in the short-wavelength limit of $|q_y|W_0 \gg 1$, the spectrum of a Tamm plasmon with the mode number $N$ tends to acquire the properties of a gated plasmon: $\omega_{Tamm, N}(q_y) = V_p \sqrt{ \pi^2 (N+1)^2 / W_0^2 + q^2_y}$.

\section{Discussion and conclusion \label{sec:DC}}

In the optical approach employed in the present investigation, as well as in other similar techniques, plasmons are treated independently, for example, in gated and ungated regions or the areas with different 2D conductivity, followed by matching the potential at the boundaries. Such methods are widely utilized to describe plasmons in various 2DESs with gates or spatial inhomogeneity of electron concentration~\cite{Satou2003, Dyakonov2008, Petrov2016, Petrov2017, Dyakonov2005, Jiang2018, rejaei2015}. However, the given technique leads to discontinuity in the plasmon potential outside the 2DES plane $z=0$ as the dependence of the potential on the $z$-coordinate changes in different regions. Nonetheless, in the case of a homogeneous 2DES, the results obtained by the optical approach are in full agreement with those produced by a more consistent method of analyzing 2DESs with single strip-shaped~\cite{zabolotnykh2019} and disk-shaped gates~\footnote{For details, see the Discussion section in Ref.~\cite{zabolotnykh2021}.}. What is more, the fact that so-called edge plasmons~\cite{Mast1985,Glattli1985,Volkov1985,Volkov1988} do not appear in the case of a homogeneous 2DES has also been established elsewhere~\cite{Petrov2016,zabolotnykh2019}, which gives additional grounds to believe that optical approach provides an accurate description of plasmons in the system. 

It should be mentioned as well that in the framework of the optical approach, the interaction of 2D electrons with their mirror-images induced in the respective gates is considered to be dominant, while the interaction between electrons in different gated region of 2DES and in different gates is regarded negligible. For this reason, the distance between the gates, $b$, is assumed to be much larger than the distance $d$ separating the 2DES and the grating.

Also, in the derivation of dispersion equations (\ref{eq:f(q)}), (\ref{eq:neq_band}), and (\ref{eq:tamm_f(q)}), we neglect the plasmon charge density excited in the ungated regions of the 2DES, taking the right-hand side of Eq.~(\ref{eq:ungated}) to be equal to zero. The charge density in ungated regions of 2DES can appear via two main reasons. The first one is the excitation of ungated plasmons, however, such plasmons are absent in the system since their frequency far exceeds that of the plasmons, that are considered in this paper, at the same wave vector $q_y$. Namely, the frequency of ungated plasmons with the wave vector $q_y$ is given as~\cite{Stern1967}:
\begin{equation}
\label{eqUng}
    \omega_p(q_y)=\sqrt{\frac{4 \pi e^2 n |q_y|}{m (\varepsilon+1)}}.
\end{equation}
By relative comparison to the characteristic frequency of the Tamm plasmon, for example, using Eq.~(\ref{eq:anal_tamm}), we find that $\omega_{Tamm}(q_y)/\omega_p(q_y) = \sqrt{2 d (\varepsilon+1)/(\varepsilon W_0)}$, which is much less than unity, as $d \ll W_0$ in our study. In general, if $\omega_p(q_y)$ is greater than the frequency of the Tamm plasmons (for instance, in the case of large $N$), it is likely that they decay into the continuum of ungated plasmons existing in the 2DES at $x<-W_0$ (Fig.~\ref{fig:system}(b)). Indeed, such decay has been considered in Ref.~\cite{zabolotnykh2021} for plasmons in a homogeneous 2DES with a disk-shaped gate. 
The second main reason of the appearance of nonzero charge density in ungated regions of 2DES is the ‘induction’ of this charge by the charge in the adjacent gate. The value of this induced in ungated region of 2DES charge can be qualitatively characterized from the estimation of the capacitance between the 2DES and the gate. Indeed, as shown, for instance, in Appendix of Ref.~\cite{Aizin2023} the capacitance has two contributions: the first one stems from the interaction of electrons in the gate with 2D electrons under the gate, while the second one corresponds to the interaction with 2D electrons in ungated area. One can see from Eq.~(A5) of Ref.~\cite{Aizin2023} that the first term is larger than the second one by the parameter $W/d$. Qualitatively, this can be understood by the fact the charge in the ungated part of 2DES is induced at distances not exceeding several $d$, see also Figs.~4(a) and (b) from Ref.~\cite{zabolotnykh2019}. Thus, in the case of our system the charge in the ungated regions of the 2DES can be neglected.

As a matter of practical significance, we list the plasmon mode numerics for the gated 2DES based on a GaAs/AlGaAs heterostructure with typical parameters as follows: the 2D electron density $n = 3 \cdot 10^{11}$ {cm}$^{-2}$, the dielectric permittivity of GaAs $\varepsilon = 12.8$, the electron effective mass $m = 0.067 m_{e}$, where $m_e$ is the mass of free electron, the distance between the gates and the 2DES $d = 20 $ nm, the gate width $W = 1$ $\mu$m, the width of the last gate $W_0 = 8 W = 8$ $\mu$m, the gate length $L = 20$ $\mu$m, and the separation distance between the gates $b = 2$ $\mu$m. Then, setting $q_y=\pi/L$ in Eq.~\eqref{eq:tamm_f(q)} results in the fundamental Tamm plasmon frequency $\omega / (2 \pi) \approx 54$ GHz and the localization length in the gated area of approximately $12$ $\mu$m (or 4 grating periods of $b + W$). It is worthwhile comparing the obtained frequency to the characteristic plasmon damping, $1/\tau$, where $\tau$ is the electron relaxation time due to the impurities and phonons. Considering GaAs/AlGaAs heterostructure at $T=4$ K, a typical value of $\tau$ can reach $40$ ps \cite{heiblum1984}. Hence, for the given frequency, it corresponds to the parameter $\omega \tau \approx 14$. Therefore, the investigated Tamm plasmons have long lifetimes and are well-defined.

Finally, let us compare briefly the obtained results with previous studies of Tamm plasmons in 2DES with grating gate performed in Ref.~\cite{Aizin2012}, in which the dispersion relation for plasmons propagating perpendicular to gates was derived, see Eq.~(23) from Ref.~\cite{Aizin2012}, via the introduction of electrical impedances for different regions of 2DES. Using local capacity approximation ($qd \ll 1$) and setting $q_2 = 0$, Eq.~(23) from Ref.~\cite{Aizin2012} becomes the same as Eq.~\eqref{eq:f(q)}, in which $q_y$ should be taken as zero. As for Tamm modes, the spectrum found in this paper at $q_y = 0$ qualitatively matches with the spectrum obtained in Ref.~\cite{Aizin2012} for the case of equal electron concentrations in gated and ungated parts of 2DES.

To conclude, we have analytically studied the plasmons propagating in a homogeneous 2DES along the grating gate formed by an array of metal strips near the 2DES, as illustrated in Fig.~\ref{fig:system}. In particular, we examine the cases of infinite and semi-infinite gratings. For the infinite grating, the plasmonic band structure is revealed and its properties are defined. In the case of semi-infinite grating, the Tamm plasmon modes can appear, localized near the edge of the grating propagating along it. The modes are shown to arise only if the last gate’s width $W_0$ differs from the width $W$ of the other gates, as indicated in Fig.~\ref{fig:system}(b). The mode frequencies are found to lie in the gaps of the plasmonic band structure. The frequencies are governed by the value of the wave vector along the strip-shaped gates, $q_y$, and the mode number, $N$, corresponding to the number of zeros of the plasmon potential under the last gate. It is demonstrated that at $W_0>W$ the spectrum of the fundamental, $N=0$, Tamm plasmon mode is in the lowest frequency gap, below the first plasmonic band. As the mode occurs only at finite values of the wave vector $q_y$, it cannot be determined analytically just by taking into account the electron dynamics across the strip-shaped gates. Indeed, the fundamental plasmon mode exists at $q_y>q_{y,0}$ and $\omega>\omega_0$, while at the point $(\omega_0,q_{y,0})$ it reaches the first plasmonic band and is no longer localized. Otherwise, at $W_0<W$ the Tamm plasmon modes lie in the higher-frequency gaps, and delocalized plasmons from the first frequency band are excitations with the lowest frequency. 

\begin{acknowledgments}
    We are grateful to Danil Rodionov and Igor Zagorodnev for valuable discussions. The work of A.A.Z. was supported by the Foundation for the Advancement of Theoretical Physics and Mathematics “BASIS” (Project No. 23-1-3-42-1).
\end{acknowledgments}

\appendix
\section{Derivation of the equation for the plasmon potential in a gated 2DES} \label{app:gated}

According to the optical approach, we consider a system of an {\it infinite} metal gate and a 2DES in the respective planes $z=d$ and $z=0$. To derive the equation for the plasmon potential in the 2DES plane $\varphi(x,z=0)$ we use Eqs.~(\ref{eq:poisson}), (\ref{eq:contin}), and (\ref{eq:cur}). To find the solution to Eq.~(\ref{eq:poisson}) we first take the Fourier transform in the $x$-direction and then solve the resultant differential equation in the $z$-direction. In that case, the boundary conditions for the potential are defined as follows: (i) it must be zero at the metal surface, $\varphi(z=d)=0$, (ii) it has to vanish at infinity, $\varphi(z \to -\infty)=0$, and (iii) it should be continuous at the 2DES plane, at $z=0$, although its derivative has a discontinuity due to the nonzero 2D charge density $\rho(q_x)$, namely, $$\partial_z \varphi(q_x,z)|^{+0}_{-0}=4 \pi \rho(q_x)/\varepsilon q,$$ where $q=\sqrt{q^2_x+q^2_y}$.

Thus, from Eq.~(\ref{eq:poisson}) we obtain the following relation:
\begin{equation}
\label{eq:app_pot}
    (1+\coth qd)\varphi(q_x,z=0)=\frac{4\pi \rho(q_x)}{\varepsilon q}.
\end{equation}

After that, we take the Fourier transform of Eqs.~(\ref{eq:contin}) and (\ref{eq:cur}) and express $\rho(q_x)$ in terms of $\varphi(q_x,z=0)$ as:
\begin{equation}
\label{eq:app_rho}
    \rho(q_x)=\sigma(\omega)q^2\varphi(q_x,z=0)/i\omega.
\end{equation}
In what follows, we use the dynamic Drude model for the conductivity, $\sigma(\omega)$, assuming the collisionless limit of $\omega\tau \gg 1$, where $\tau$ is the electron relaxation time. This leads to $\sigma(\omega)=e^2 n/(-i\omega m)$ with $-e$, $m$, and $n$ denoting the electron charge, the effective mass, and the 2D concentration. Then, the substitution of $\rho(q_x)$ from Eq.~(\ref{eq:app_rho}) into Eq.~(\ref{eq:app_pot}) yields:   
\begin{equation}
\label{eq:app_pot2}
    \left(1+\coth(qd) -\frac{4\pi e^2 n q}{\varepsilon m \omega^2}\right)\varphi(q_x,z=0)=0.
\end{equation}
Next, in the long-wavelength limit of $qd \ll 1$, Eq.~(\ref{eq:app_pot2}) can be expressed as:
\begin{equation}
\label{eq:app_pot3}
    \left( q_x^2+q_y^2-\frac{\omega^2}{V_p^2}  \right)\varphi(q_x,z=0)=0,
\end{equation}
where $V_p=\sqrt{4\pi e^2 n d/\varepsilon m}$ is the velocity of plasmons in the gated 2DES in the long-wavelength limit.

Finally, taking the inverse Fourier transform of Eq.~(\ref{eq:app_pot3}) we arrive at Eq.~(\ref{eq:gated}) as desired.

\section{Derivation of the equations for infinite and semi-infinite grating gates} \label{app:lambdas}

Given that plasmon potential in gated and ungated regions of the 2DES is described by Eqs.~(\ref{eq:gated}) and (\ref{eq:ungated}), respectively, their solutions can be written as follows:
\begin{align}
    \label{eq:sol_gat}
    &\varphi(x_l) = A_{l} e^{ik(x_l - b - l (b + W))} + B_{l} e^{-ik(x_l - b - l (b + W))},\\
    \label{eq:sol_ungat}
    &\varphi(x'_l) =  C_{l} e^{|q_y|(x'_l - l (b + W))} + D_{l} e^{-|q_y|(x'_l - l (b + W))},
\end{align}
where $k=\sqrt{\omega^2/V_p^2- q_y^2}$, while the coordinates $x_l$ and $x'_l$ refer to the 2DES regions under the $l$-th gate and between the $(l-1)$-th and the $l$-th gates, accordingly. To determine the plasmon potential in a 2DES with the grating, Eqs.~(\ref{eq:sol_gat}) and (\ref{eq:sol_ungat}) should be solved for the boundary conditions of the continuity of the potential and its $x$-derivative.

To begin with, we look for $A_l$, $B_l$, $C_l$, $D_l$ in the form $(A_l, B_l, C_l, D_l) = \lambda^l (A_0, B_0, C_0, D_0)$, where the constant $\lambda$ is a 'scaling' factor. 
 
Consequently, for the given boundary conditions, we obtain the following set of equations: 
\begin{widetext}
\begin{equation}
\label{Mat}
\begin{pmatrix}
    e^{ikb} & e^{-ikb} & - e^{|q_y| b} & - e^{-|q_y| b}\\
    ik e^{ikb} & -ik e^{-ikb} & -|q_y| e^{|q_y| b} & |q_y| e^{-|q_y| b}\\
    e^{ik(b + W)} & e^{-ik(b + W)} & - \lambda & -\lambda\\
    ik e^{ik(b + W)} & - ik e^{-ik(b + W)} & -\lambda |q_y| & \lambda |q_y|
\end{pmatrix}
\begin{pmatrix}
    A_l\\
    B_l\\
    C_l\\
    D_l
\end{pmatrix}
=0.
\end{equation}
\end{widetext}

The dispersion equation for plasmons in the 2DES with infinite grating can be found by setting the determinant of the matrix in the left-hand side of~(\ref{Mat}) to zero, which implies the existence of a non-trivial solution of the set~(\ref{Mat}). Thus, the dispersion relation becomes: 
\begin{equation}
    \label{eq:lambda}
    \lambda^2 - 2 \lambda f(q_y, \omega)  + 1 = 0,
\end{equation}
where $f(q_y, \omega)$ is defined according to~(\ref{eq:f(q)}). Here, we note that for the solutions to exist, the potential should be finite, i.e. $|\lambda| = 1$, which leads to the condition of $-1 \le f(q_y,\omega) \le 1$ and, ultimately, to the formation of the band structure for the plasmon frequencies.

As for the 2DES with semi-infinite grating, Fig.~\ref{fig:system}(b), the plasmon potential at $x < 0$ can be formulated as:
\begin{align}
    \label{eq:phi_ung_neg}
    &\varphi(x) = C_{-1}e^{|q_y|x}, \ \  x < - W_0,\\
    \label{eq:phi_gat_neg}
    &\varphi(x) = A_{-1} e^{ikx} +B_{-1} e ^{-ikx},  -W_0 \leq x < 0.
\end{align}

Otherwise, at $x>0$, the potential can be determined following the same procedure used for the infinite grating, by introducing the parameter $\lambda$ and the relation between the coefficients~(\ref{Mat}).

Hence, based on Eqs.~(\ref{Mat}), (\ref{eq:phi_ung_neg}), and (\ref{eq:phi_gat_neg}) as well as the boundary conditions of continuity of $\varphi(x)$ and its $x$-derivative at $x=0$ and $x = -W_0$ and the condition of vanishing $\varphi(x)$ at $x \to \pm \infty$, the dispersion equation for the Tamm plasmon is derived to be:
\begin{equation}
    \label{eq:almost_disp}
    \frac{q_y^2 + k^2}{q_y^2 - k^2 + 2|q_y|k \cot{kW_0}} = \frac{|q_y| + i k}{|q_y| - i k} \frac{\lambda - e^{ikW + |q_y|b}}{\lambda - e^{ikW - |q_y|b}}.
\end{equation}

We note that the left-hand side of Eq.~\eqref{eq:almost_disp} is purely real, whereas the right-hand side is complex-valued, with the imaginary part given by:
$$ \frac{\lambda^2 - 2 \lambda f(q_y, \omega)  + 1}
    {2 |q_y| k(q_y^2 + k^2) \left(\lambda^2 - 2\lambda e^{-|q_y|b}\cos{kW} + e^{-2|q_y|b} \right)},$$
which equals zero according to Eq.~\eqref{eq:lambda}. At the same time, the real part can be expressed as: $$\frac{q_y^2 - k^2}{q_y^2 + k^2} 
\frac{\lambda^2 - 2 \lambda f(q_y, \omega) + 1 + \lambda \frac{(q_y^2 + k^2)^2 \sin{kW} \sinh{|q_y|b}}{|q_y| k(q_y^2 - k^2) }}{\lambda^2 - 2\lambda e^{-|q_y|b}\cos{kW} + e^{-2|q_y|b}}. $$

Therefore, based on Eq.~\eqref{eq:lambda}, the dispersion relation for the Tamm plasmon can be written in terms of factor $\lambda_{Tamm}$ as follows: 
\begin{equation}
    \label{eq:disp_tamm_la}
    \lambda_{Tamm} = e^{-|q_y|b} \frac{\sin{kW_0}}{\sin{k(W_0 - W)}}.
\end{equation}
At last, applying the expression for $\lambda_{Tamm}$ from Eq.~\eqref{eq:lambda} to the derived Eq.~\eqref{eq:disp_tamm_la}, we arrive at the desired Eq.~\eqref{eq:tamm_f(q)}.

\section{Plasmon spectra in 2DES with a semi-infinite grating at $W_0<W$} \label{AppC}
\begin{figure*}
    \includegraphics[width=2.0\columnwidth]{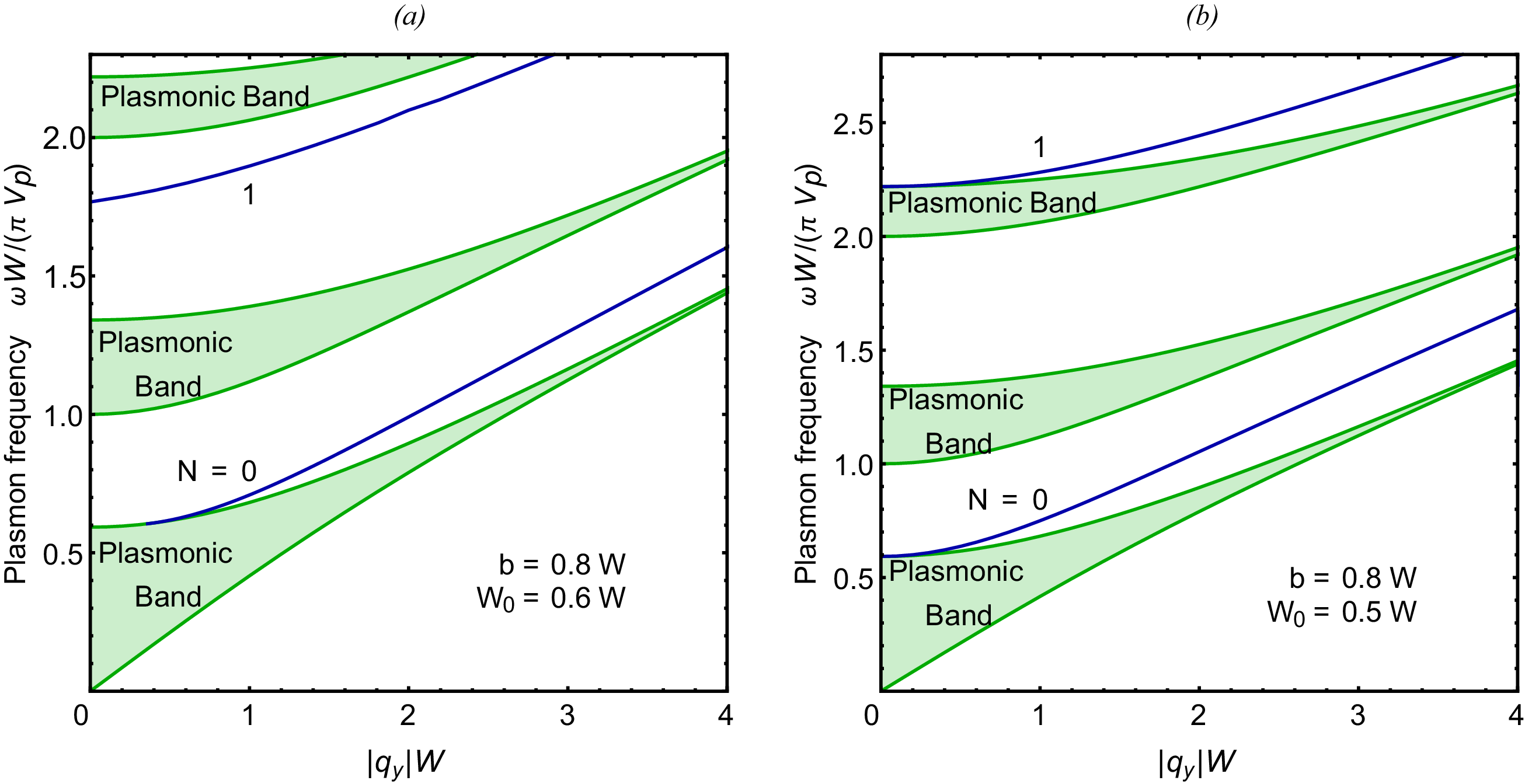}
    \caption{\label{fig:Spec2}Spectra of plasmons in the 2DES with a semi-infinite grating gate, as defined by Eqs.~\eqref{eq:neq_band} and \eqref{eq:tamm_f(q)}. The regions shaded in light green represent the plasmon frequency bands. The blue lines denote the Tamm plasmon modes $N = 0, 1$. Parameters $b / W$ and $W_0 / W$ are set to 0.8 and 0.6 in Fig.~\ref{fig:Spec2} $(a)$, 0.8 and 0.5 in Fig.~\ref{fig:Spec2} $(b)$ respectively.}
\end{figure*}
Plasmon spectra for the case of $W_0 < W$ are shown in Fig.~\ref{fig:Spec2}. In this case, the fundamental Tamm mode, $N=0$, lies in the second frequency gap. Plasmon spectrum in the system with $W_0=W/2$, when Tamm plasmons exist only in even plasmonic gaps, is depicted in Fig.~\ref{fig:Spec2}(b), for details see the paragraph before and after Eq.~(\ref{eq:tamm_f(q)_0.5}) in the main text.

\bibliography{bibl}

\begin{thebibliography}{67}%
\makeatletter
\providecommand \@ifxundefined [1]{%
 \@ifx{#1\undefined}
}%
\providecommand \@ifnum [1]{%
 \ifnum #1\expandafter \@firstoftwo
 \else \expandafter \@secondoftwo
 \fi
}%
\providecommand \@ifx [1]{%
 \ifx #1\expandafter \@firstoftwo
 \else \expandafter \@secondoftwo
 \fi
}%
\providecommand \natexlab [1]{#1}%
\providecommand \enquote  [1]{``#1''}%
\providecommand \bibnamefont  [1]{#1}%
\providecommand \bibfnamefont [1]{#1}%
\providecommand \citenamefont [1]{#1}%
\providecommand \href@noop [0]{\@secondoftwo}%
\providecommand \href [0]{\begingroup \@sanitize@url \@href}%
\providecommand \@href[1]{\@@startlink{#1}\@@href}%
\providecommand \@@href[1]{\endgroup#1\@@endlink}%
\providecommand \@sanitize@url [0]{\catcode `\\12\catcode `\$12\catcode `\&12\catcode `\#12\catcode `\^12\catcode `\_12\catcode `\%12\relax}%
\providecommand \@@startlink[1]{}%
\providecommand \@@endlink[0]{}%
\providecommand \url  [0]{\begingroup\@sanitize@url \@url }%
\providecommand \@url [1]{\endgroup\@href {#1}{\urlprefix }}%
\providecommand \urlprefix  [0]{URL }%
\providecommand \Eprint [0]{\href }%
\providecommand \doibase [0]{https://doi.org/}%
\providecommand \selectlanguage [0]{\@gobble}%
\providecommand \bibinfo  [0]{\@secondoftwo}%
\providecommand \bibfield  [0]{\@secondoftwo}%
\providecommand \translation [1]{[#1]}%
\providecommand \BibitemOpen [0]{}%
\providecommand \bibitemStop [0]{}%
\providecommand \bibitemNoStop [0]{.\EOS\space}%
\providecommand \EOS [0]{\spacefactor3000\relax}%
\providecommand \BibitemShut  [1]{\csname bibitem#1\endcsname}%
\let\auto@bib@innerbib\@empty
\bibitem [{\citenamefont {Stern}(1967)}]{Stern1967}%
  \BibitemOpen
  \bibfield  {author} {\bibinfo {author} {\bibfnamefont {F.}~\bibnamefont {Stern}},\ }\bibfield  {title} {\bibinfo {title} {Polarizability of a two-dimensional electron gas},\ }\href {https://doi.org/10.1103/PhysRevLett.18.546} {\bibfield  {journal} {\bibinfo  {journal} {Phys. Rev. Lett.}\ }\textbf {\bibinfo {volume} {18}},\ \bibinfo {pages} {546} (\bibinfo {year} {1967})}\BibitemShut {NoStop}%
\bibitem [{\citenamefont {Chaplik}(1972)}]{chaplik1972}%
  \BibitemOpen
  \bibfield  {author} {\bibinfo {author} {\bibfnamefont {A.~V.}\ \bibnamefont {Chaplik}},\ }\bibfield  {title} {\bibinfo {title} {Possible crystallization of charge carriers in low-density inversion layers},\ }\href {http://www.jetp.ras.ru/cgi-bin/dn/e_035_02_0395.pdf} {\bibfield  {journal} {\bibinfo  {journal} {Zh. Eksp. Teor. Fiz.}\ }\textbf {\bibinfo {volume} {62}},\ \bibinfo {pages} {746} (\bibinfo {year} {1972})},\ \translation{Sov. Phys. JETP \textbf{35}, 395 (1972)}\BibitemShut {NoStop}%
\bibitem [{\citenamefont {Grimes}\ and\ \citenamefont {Adams}(1976)}]{Grimes1976}%
  \BibitemOpen
  \bibfield  {author} {\bibinfo {author} {\bibfnamefont {C.~C.}\ \bibnamefont {Grimes}}\ and\ \bibinfo {author} {\bibfnamefont {G.}~\bibnamefont {Adams}},\ }\bibfield  {title} {\bibinfo {title} {Observation of two-dimensional plasmons and electron-ripplon scattering in a sheet of electrons on liquid helium},\ }\href {https://doi.org/10.1103/PhysRevLett.36.145} {\bibfield  {journal} {\bibinfo  {journal} {Phys. Rev. Lett.}\ }\textbf {\bibinfo {volume} {36}},\ \bibinfo {pages} {145} (\bibinfo {year} {1976})}\BibitemShut {NoStop}%
\bibitem [{\citenamefont {Allen~Jr.}\ \emph {et~al.}(1977)\citenamefont {Allen~Jr.}, \citenamefont {Tsui},\ and\ \citenamefont {Logan}}]{Allen1977}%
  \BibitemOpen
  \bibfield  {author} {\bibinfo {author} {\bibfnamefont {S.~J.}\ \bibnamefont {Allen~Jr.}}, \bibinfo {author} {\bibfnamefont {D.~C.}\ \bibnamefont {Tsui}},\ and\ \bibinfo {author} {\bibfnamefont {R.~A.}\ \bibnamefont {Logan}},\ }\bibfield  {title} {\bibinfo {title} {Observation of the two-dimensional plasmon in silicon inversion layers},\ }\href {https://doi.org/10.1103/PhysRevLett.38.980} {\bibfield  {journal} {\bibinfo  {journal} {Phys. Rev. Lett.}\ }\textbf {\bibinfo {volume} {38}},\ \bibinfo {pages} {980} (\bibinfo {year} {1977})}\BibitemShut {NoStop}%
\bibitem [{\citenamefont {Theis}\ \emph {et~al.}(1977)\citenamefont {Theis}, \citenamefont {Kotthaus},\ and\ \citenamefont {Stiles}}]{Theis1977}%
  \BibitemOpen
  \bibfield  {author} {\bibinfo {author} {\bibfnamefont {T.~N.}\ \bibnamefont {Theis}}, \bibinfo {author} {\bibfnamefont {J.~P.}\ \bibnamefont {Kotthaus}},\ and\ \bibinfo {author} {\bibfnamefont {P.~J.}\ \bibnamefont {Stiles}},\ }\bibfield  {title} {\bibinfo {title} {Two-dimensional magnetoplasmon in the silicon inversion layer},\ }\href {https://doi.org/10.1016/0038-1098(77)90205-8} {\bibfield  {journal} {\bibinfo  {journal} {Solid State Commun.}\ }\textbf {\bibinfo {volume} {24}},\ \bibinfo {pages} {273} (\bibinfo {year} {1977})}\BibitemShut {NoStop}%
\bibitem [{\citenamefont {Kukushkin}\ \emph {et~al.}(2003)\citenamefont {Kukushkin}, \citenamefont {Smet}, \citenamefont {Mikhailov}, \citenamefont {Kulakovskii}, \citenamefont {von Klitzing},\ and\ \citenamefont {Wegscheider}}]{Kukushkin2003}%
  \BibitemOpen
  \bibfield  {author} {\bibinfo {author} {\bibfnamefont {I.~V.}\ \bibnamefont {Kukushkin}}, \bibinfo {author} {\bibfnamefont {J.~H.}\ \bibnamefont {Smet}}, \bibinfo {author} {\bibfnamefont {S.~A.}\ \bibnamefont {Mikhailov}}, \bibinfo {author} {\bibfnamefont {D.~V.}\ \bibnamefont {Kulakovskii}}, \bibinfo {author} {\bibfnamefont {K.}~\bibnamefont {von Klitzing}},\ and\ \bibinfo {author} {\bibfnamefont {W.}~\bibnamefont {Wegscheider}},\ }\bibfield  {title} {\bibinfo {title} {Observation of retardation effects in the spectrum of two-dimensional plasmons},\ }\href {https://doi.org/10.1103/PhysRevLett.90.156801} {\bibfield  {journal} {\bibinfo  {journal} {Phys. Rev. Lett.}\ }\textbf {\bibinfo {volume} {90}},\ \bibinfo {pages} {156801} (\bibinfo {year} {2003})}\BibitemShut {NoStop}%
\bibitem [{\citenamefont {Muravev}\ \emph {et~al.}(2008)\citenamefont {Muravev}, \citenamefont {Fortunatov}, \citenamefont {Kukushkin}, \citenamefont {Smet}, \citenamefont {Dietsche},\ and\ \citenamefont {von Klitzing}}]{Muravev2008}%
  \BibitemOpen
  \bibfield  {author} {\bibinfo {author} {\bibfnamefont {V.~M.}\ \bibnamefont {Muravev}}, \bibinfo {author} {\bibfnamefont {A.~A.}\ \bibnamefont {Fortunatov}}, \bibinfo {author} {\bibfnamefont {I.~V.}\ \bibnamefont {Kukushkin}}, \bibinfo {author} {\bibfnamefont {J.~H.}\ \bibnamefont {Smet}}, \bibinfo {author} {\bibfnamefont {W.}~\bibnamefont {Dietsche}},\ and\ \bibinfo {author} {\bibfnamefont {K.}~\bibnamefont {von Klitzing}},\ }\bibfield  {title} {\bibinfo {title} {Tunable plasmonic crystals for edge magnetoplasmons of a two-dimensional electron system},\ }\href {https://doi.org/10.1103/PhysRevLett.101.216801} {\bibfield  {journal} {\bibinfo  {journal} {Phys. Rev. Lett.}\ }\textbf {\bibinfo {volume} {101}},\ \bibinfo {pages} {216801} (\bibinfo {year} {2008})}\BibitemShut {NoStop}%
\bibitem [{\citenamefont {Muravev}\ \emph {et~al.}(2011)\citenamefont {Muravev}, \citenamefont {Andreev}, \citenamefont {Kukushkin}, \citenamefont {Schmult},\ and\ \citenamefont {Dietsche}}]{Muravev2011}%
  \BibitemOpen
  \bibfield  {author} {\bibinfo {author} {\bibfnamefont {V.~M.}\ \bibnamefont {Muravev}}, \bibinfo {author} {\bibfnamefont {I.~V.}\ \bibnamefont {Andreev}}, \bibinfo {author} {\bibfnamefont {I.~V.}\ \bibnamefont {Kukushkin}}, \bibinfo {author} {\bibfnamefont {S.}~\bibnamefont {Schmult}},\ and\ \bibinfo {author} {\bibfnamefont {W.}~\bibnamefont {Dietsche}},\ }\bibfield  {title} {\bibinfo {title} {Observation of hybrid plasmon-photon modes in microwave transmission of coplanar microresonators},\ }\href {https://doi.org/10.1103/PhysRevB.83.075309} {\bibfield  {journal} {\bibinfo  {journal} {Phys. Rev. B}\ }\textbf {\bibinfo {volume} {83}},\ \bibinfo {pages} {075309} (\bibinfo {year} {2011})}\BibitemShut {NoStop}%
\bibitem [{\citenamefont {Scalari}\ \emph {et~al.}(2012)\citenamefont {Scalari}, \citenamefont {Maissen}, \citenamefont {Tur{\v{c}}inkov{\'a}}, \citenamefont {Hagenm{\"u}ller}, \citenamefont {De~Liberato}, \citenamefont {Ciuti}, \citenamefont {Reichl}, \citenamefont {Schuh}, \citenamefont {Wegscheider}, \citenamefont {Beck},\ and\ \citenamefont {Faist}}]{scalari2012}%
  \BibitemOpen
  \bibfield  {author} {\bibinfo {author} {\bibfnamefont {G.}~\bibnamefont {Scalari}}, \bibinfo {author} {\bibfnamefont {C.}~\bibnamefont {Maissen}}, \bibinfo {author} {\bibfnamefont {D.}~\bibnamefont {Tur{\v{c}}inkov{\'a}}}, \bibinfo {author} {\bibfnamefont {D.}~\bibnamefont {Hagenm{\"u}ller}}, \bibinfo {author} {\bibfnamefont {S.}~\bibnamefont {De~Liberato}}, \bibinfo {author} {\bibfnamefont {C.}~\bibnamefont {Ciuti}}, \bibinfo {author} {\bibfnamefont {C.}~\bibnamefont {Reichl}}, \bibinfo {author} {\bibfnamefont {D.}~\bibnamefont {Schuh}}, \bibinfo {author} {\bibfnamefont {W.}~\bibnamefont {Wegscheider}}, \bibinfo {author} {\bibfnamefont {M.}~\bibnamefont {Beck}},\ and\ \bibinfo {author} {\bibfnamefont {J.}~\bibnamefont {Faist}},\ }\bibfield  {title} {\bibinfo {title} {Ultrastrong coupling of the cyclotron transition of a 2{D} electron gas to a {TH}z metamaterial},\ }\href {https://doi.org/10.1126/science.1216022} {\bibfield  {journal} {\bibinfo  {journal} {Science}\ }\textbf {\bibinfo {volume} {335}},\
  \bibinfo {pages} {1323} (\bibinfo {year} {2012})}\BibitemShut {NoStop}%
\bibitem [{\citenamefont {Dyer}\ \emph {et~al.}(2013)\citenamefont {Dyer}, \citenamefont {Aizin}, \citenamefont {Allen}, \citenamefont {Grine}, \citenamefont {Bethke}, \citenamefont {Reno},\ and\ \citenamefont {Shaner}}]{Dyer2013}%
  \BibitemOpen
  \bibfield  {author} {\bibinfo {author} {\bibfnamefont {G.~C.}\ \bibnamefont {Dyer}}, \bibinfo {author} {\bibfnamefont {G.~R.}\ \bibnamefont {Aizin}}, \bibinfo {author} {\bibfnamefont {S.~J.}\ \bibnamefont {Allen}}, \bibinfo {author} {\bibfnamefont {A.~D.}\ \bibnamefont {Grine}}, \bibinfo {author} {\bibfnamefont {D.}~\bibnamefont {Bethke}}, \bibinfo {author} {\bibfnamefont {J.~L.}\ \bibnamefont {Reno}},\ and\ \bibinfo {author} {\bibfnamefont {E.~A.}\ \bibnamefont {Shaner}},\ }\bibfield  {title} {\bibinfo {title} {Induced transparency by coupling of {T}amm and defect states in tunable terahertz plasmonic crystals},\ }\href {https://doi.org/10.1038/nphoton.2013.252} {\bibfield  {journal} {\bibinfo  {journal} {Nat. Photonics}\ }\textbf {\bibinfo {volume} {7}},\ \bibinfo {pages} {925} (\bibinfo {year} {2013})}\BibitemShut {NoStop}%
\bibitem [{\citenamefont {\L{}usakowski}(2016)}]{Lusakowski2016}%
  \BibitemOpen
  \bibfield  {author} {\bibinfo {author} {\bibfnamefont {J.}~\bibnamefont {\L{}usakowski}},\ }\bibfield  {title} {\bibinfo {title} {Plasmon--terahertz photon interaction in high-electron-mobility heterostructures},\ }\href {https://doi.org/10.1088/0268-1242/32/1/013004} {\bibfield  {journal} {\bibinfo  {journal} {Semicond. Sci. Technol.}\ }\textbf {\bibinfo {volume} {32}},\ \bibinfo {pages} {013004} (\bibinfo {year} {2016})}\BibitemShut {NoStop}%
\bibitem [{\citenamefont {Muravev}\ \emph {et~al.}(2015)\citenamefont {Muravev}, \citenamefont {Gusikhin}, \citenamefont {Andreev},\ and\ \citenamefont {Kukushkin}}]{muravev2015novel}%
  \BibitemOpen
  \bibfield  {author} {\bibinfo {author} {\bibfnamefont {V.~M.}\ \bibnamefont {Muravev}}, \bibinfo {author} {\bibfnamefont {P.~A.}\ \bibnamefont {Gusikhin}}, \bibinfo {author} {\bibfnamefont {I.~V.}\ \bibnamefont {Andreev}},\ and\ \bibinfo {author} {\bibfnamefont {I.~V.}\ \bibnamefont {Kukushkin}},\ }\bibfield  {title} {\bibinfo {title} {Novel relativistic plasma excitations in a gated two-dimensional electron system},\ }\href {https://doi.org/10.1103/PhysRevLett.114.106805} {\bibfield  {journal} {\bibinfo  {journal} {Phys. Rev. Lett.}\ }\textbf {\bibinfo {volume} {114}},\ \bibinfo {pages} {106805} (\bibinfo {year} {2015})}\BibitemShut {NoStop}%
\bibitem [{\citenamefont {Muravev}\ and\ \citenamefont {Kukushkin}(2020)}]{muravev2020collective}%
  \BibitemOpen
  \bibfield  {author} {\bibinfo {author} {\bibfnamefont {V.~M.}\ \bibnamefont {Muravev}}\ and\ \bibinfo {author} {\bibfnamefont {I.~V.}\ \bibnamefont {Kukushkin}},\ }\bibfield  {title} {\bibinfo {title} {Collective plasma excitations in two-dimensional electron systems},\ }\href {https://doi.org/10.3367/UFNe.2019.07.038637} {\bibfield  {journal} {\bibinfo  {journal} {Phys.-Usp.}\ }\textbf {\bibinfo {volume} {63}},\ \bibinfo {pages} {975} (\bibinfo {year} {2020})}\BibitemShut {NoStop}%
\bibitem [{\citenamefont {Shuvaev}\ \emph {et~al.}(2021)\citenamefont {Shuvaev}, \citenamefont {Muravev}, \citenamefont {Gusikhin}, \citenamefont {Gospodari{\v{c}}}, \citenamefont {Pimenov},\ and\ \citenamefont {Kukushkin}}]{Shuvaev2021}%
  \BibitemOpen
  \bibfield  {author} {\bibinfo {author} {\bibfnamefont {A.}~\bibnamefont {Shuvaev}}, \bibinfo {author} {\bibfnamefont {V.~M.}\ \bibnamefont {Muravev}}, \bibinfo {author} {\bibfnamefont {P.~A.}\ \bibnamefont {Gusikhin}}, \bibinfo {author} {\bibfnamefont {J.}~\bibnamefont {Gospodari{\v{c}}}}, \bibinfo {author} {\bibfnamefont {A.}~\bibnamefont {Pimenov}},\ and\ \bibinfo {author} {\bibfnamefont {I.~V.}\ \bibnamefont {Kukushkin}},\ }\bibfield  {title} {\bibinfo {title} {Discovery of two-dimensional electromagnetic plasma waves},\ }\href {https://doi.org/10.1103/PhysRevLett.126.136801} {\bibfield  {journal} {\bibinfo  {journal} {Phys. Rev. Lett.}\ }\textbf {\bibinfo {volume} {126}},\ \bibinfo {pages} {136801} (\bibinfo {year} {2021})}\BibitemShut {NoStop}%
\bibitem [{\citenamefont {Andreev}\ \emph {et~al.}(2021)\citenamefont {Andreev}, \citenamefont {Muravev}, \citenamefont {Semenov},\ and\ \citenamefont {Kukushkin}}]{Andreev2021}%
  \BibitemOpen
  \bibfield  {author} {\bibinfo {author} {\bibfnamefont {I.~V.}\ \bibnamefont {Andreev}}, \bibinfo {author} {\bibfnamefont {V.~M.}\ \bibnamefont {Muravev}}, \bibinfo {author} {\bibfnamefont {N.~D.}\ \bibnamefont {Semenov}},\ and\ \bibinfo {author} {\bibfnamefont {I.~V.}\ \bibnamefont {Kukushkin}},\ }\bibfield  {title} {\bibinfo {title} {Observation of acoustic plasma waves with a velocity approaching the speed of light},\ }\href {https://doi.org/10.1103/PhysRevB.103.115420} {\bibfield  {journal} {\bibinfo  {journal} {Phys. Rev. B}\ }\textbf {\bibinfo {volume} {103}},\ \bibinfo {pages} {115420} (\bibinfo {year} {2021})}\BibitemShut {NoStop}%
\bibitem [{\citenamefont {Woessner}\ \emph {et~al.}(2015)\citenamefont {Woessner}, \citenamefont {Lundeberg}, \citenamefont {Gao}, \citenamefont {Principi}, \citenamefont {Alonso-Gonz{\'a}lez}, \citenamefont {Carrega}, \citenamefont {Watanabe}, \citenamefont {Taniguchi}, \citenamefont {Vignale}, \citenamefont {Polini}, \citenamefont {Hillenbrand},\ and\ \citenamefont {Koppens}}]{woessner2015}%
  \BibitemOpen
  \bibfield  {author} {\bibinfo {author} {\bibfnamefont {A.}~\bibnamefont {Woessner}}, \bibinfo {author} {\bibfnamefont {M.~B.}\ \bibnamefont {Lundeberg}}, \bibinfo {author} {\bibfnamefont {Y.}~\bibnamefont {Gao}}, \bibinfo {author} {\bibfnamefont {A.}~\bibnamefont {Principi}}, \bibinfo {author} {\bibfnamefont {P.}~\bibnamefont {Alonso-Gonz{\'a}lez}}, \bibinfo {author} {\bibfnamefont {M.}~\bibnamefont {Carrega}}, \bibinfo {author} {\bibfnamefont {K.}~\bibnamefont {Watanabe}}, \bibinfo {author} {\bibfnamefont {T.}~\bibnamefont {Taniguchi}}, \bibinfo {author} {\bibfnamefont {G.}~\bibnamefont {Vignale}}, \bibinfo {author} {\bibfnamefont {M.}~\bibnamefont {Polini}}, \bibinfo {author} {\bibfnamefont {R.}~\bibnamefont {Hillenbrand}},\ and\ \bibinfo {author} {\bibfnamefont {F.~H.~L.}\ \bibnamefont {Koppens}},\ }\bibfield  {title} {\bibinfo {title} {Highly confined low-loss plasmons in graphene--boron nitride heterostructures},\ }\href {https://doi.org/10.1038/nmat4169} {\bibfield  {journal} {\bibinfo  {journal}
  {Nat. Mater.}\ }\textbf {\bibinfo {volume} {14}},\ \bibinfo {pages} {421} (\bibinfo {year} {2015})}\BibitemShut {NoStop}%
\bibitem [{\citenamefont {Iranzo}\ \emph {et~al.}(2018)\citenamefont {Iranzo}, \citenamefont {Nanot}, \citenamefont {Dias}, \citenamefont {Epstein}, \citenamefont {Peng}, \citenamefont {Efetov}, \citenamefont {Lundeberg}, \citenamefont {Parret}, \citenamefont {Osmond}, \citenamefont {Hong}, \citenamefont {Kong}, \citenamefont {Englund}, \citenamefont {Peres},\ and\ \citenamefont {Koppens}}]{Iranzo2018}%
  \BibitemOpen
  \bibfield  {author} {\bibinfo {author} {\bibfnamefont {D.~A.}\ \bibnamefont {Iranzo}}, \bibinfo {author} {\bibfnamefont {S.}~\bibnamefont {Nanot}}, \bibinfo {author} {\bibfnamefont {E.~J.~C.}\ \bibnamefont {Dias}}, \bibinfo {author} {\bibfnamefont {I.}~\bibnamefont {Epstein}}, \bibinfo {author} {\bibfnamefont {C.}~\bibnamefont {Peng}}, \bibinfo {author} {\bibfnamefont {D.~K.}\ \bibnamefont {Efetov}}, \bibinfo {author} {\bibfnamefont {M.~B.}\ \bibnamefont {Lundeberg}}, \bibinfo {author} {\bibfnamefont {R.}~\bibnamefont {Parret}}, \bibinfo {author} {\bibfnamefont {J.}~\bibnamefont {Osmond}}, \bibinfo {author} {\bibfnamefont {J.-Y.}\ \bibnamefont {Hong}}, \bibinfo {author} {\bibfnamefont {J.}~\bibnamefont {Kong}}, \bibinfo {author} {\bibfnamefont {D.~R.}\ \bibnamefont {Englund}}, \bibinfo {author} {\bibfnamefont {N.~M.~R.}\ \bibnamefont {Peres}},\ and\ \bibinfo {author} {\bibfnamefont {F.~H.~L.}\ \bibnamefont {Koppens}},\ }\bibfield  {title} {\bibinfo {title} {Probing the ultimate plasmon confinement limits
  with a van der {W}aals heterostructure},\ }\href {https://doi.org/10.1126/science.aar8438} {\bibfield  {journal} {\bibinfo  {journal} {Science}\ }\textbf {\bibinfo {volume} {360}},\ \bibinfo {pages} {291} (\bibinfo {year} {2018})}\BibitemShut {NoStop}%
\bibitem [{\citenamefont {Bandurin}\ \emph {et~al.}(2018)\citenamefont {Bandurin}, \citenamefont {Svintsov}, \citenamefont {Gayduchenko}, \citenamefont {Xu}, \citenamefont {Principi}, \citenamefont {Moskotin}, \citenamefont {Tretyakov}, \citenamefont {Yagodkin}, \citenamefont {Zhukov}, \citenamefont {Taniguchi}, \citenamefont {Watanabe}, \citenamefont {Grigorieva}, \citenamefont {Polini}, \citenamefont {Goltsman}, \citenamefont {Geim},\ and\ \citenamefont {Fedorov}}]{Bandurin2018}%
  \BibitemOpen
  \bibfield  {author} {\bibinfo {author} {\bibfnamefont {D.~A.}\ \bibnamefont {Bandurin}}, \bibinfo {author} {\bibfnamefont {D.}~\bibnamefont {Svintsov}}, \bibinfo {author} {\bibfnamefont {I.}~\bibnamefont {Gayduchenko}}, \bibinfo {author} {\bibfnamefont {S.~G.}\ \bibnamefont {Xu}}, \bibinfo {author} {\bibfnamefont {A.}~\bibnamefont {Principi}}, \bibinfo {author} {\bibfnamefont {M.}~\bibnamefont {Moskotin}}, \bibinfo {author} {\bibfnamefont {I.}~\bibnamefont {Tretyakov}}, \bibinfo {author} {\bibfnamefont {D.}~\bibnamefont {Yagodkin}}, \bibinfo {author} {\bibfnamefont {S.}~\bibnamefont {Zhukov}}, \bibinfo {author} {\bibfnamefont {T.}~\bibnamefont {Taniguchi}}, \bibinfo {author} {\bibfnamefont {K.}~\bibnamefont {Watanabe}}, \bibinfo {author} {\bibfnamefont {I.~V.}\ \bibnamefont {Grigorieva}}, \bibinfo {author} {\bibfnamefont {M.}~\bibnamefont {Polini}}, \bibinfo {author} {\bibfnamefont {G.}~\bibnamefont {Goltsman}}, \bibinfo {author} {\bibfnamefont {A.~K.}\ \bibnamefont {Geim}},\ and\ \bibinfo {author}
  {\bibfnamefont {G.}~\bibnamefont {Fedorov}},\ }\bibfield  {title} {\bibinfo {title} {Resonant terahertz detection using graphene plasmons},\ }\href {https://doi.org/10.1038/s41467-018-07848-w} {\bibfield  {journal} {\bibinfo  {journal} {Nat. Commun.}\ }\textbf {\bibinfo {volume} {9}},\ \bibinfo {pages} {1} (\bibinfo {year} {2018})}\BibitemShut {NoStop}%
\bibitem [{\citenamefont {Bylinkin}\ \emph {et~al.}(2019)\citenamefont {Bylinkin}, \citenamefont {Titova}, \citenamefont {Mikheev}, \citenamefont {Zhukova}, \citenamefont {Zhukov}, \citenamefont {Belyanchikov}, \citenamefont {Kashchenko}, \citenamefont {Miakonkikh},\ and\ \citenamefont {Svintsov}}]{Bylinkin2019}%
  \BibitemOpen
  \bibfield  {author} {\bibinfo {author} {\bibfnamefont {A.}~\bibnamefont {Bylinkin}}, \bibinfo {author} {\bibfnamefont {E.}~\bibnamefont {Titova}}, \bibinfo {author} {\bibfnamefont {V.}~\bibnamefont {Mikheev}}, \bibinfo {author} {\bibfnamefont {E.}~\bibnamefont {Zhukova}}, \bibinfo {author} {\bibfnamefont {S.}~\bibnamefont {Zhukov}}, \bibinfo {author} {\bibfnamefont {M.}~\bibnamefont {Belyanchikov}}, \bibinfo {author} {\bibfnamefont {M.}~\bibnamefont {Kashchenko}}, \bibinfo {author} {\bibfnamefont {A.}~\bibnamefont {Miakonkikh}},\ and\ \bibinfo {author} {\bibfnamefont {D.}~\bibnamefont {Svintsov}},\ }\bibfield  {title} {\bibinfo {title} {Tight-binding terahertz plasmons in chemical-vapor-deposited graphene},\ }\href {https://doi.org/10.1103/PhysRevApplied.11.054017} {\bibfield  {journal} {\bibinfo  {journal} {Phys. Rev. Applied}\ }\textbf {\bibinfo {volume} {11}},\ \bibinfo {pages} {054017} (\bibinfo {year} {2019})}\BibitemShut {NoStop}%
\bibitem [{\citenamefont {Epstein}\ \emph {et~al.}(2020)\citenamefont {Epstein}, \citenamefont {Alcaraz}, \citenamefont {Huang}, \citenamefont {Pusapati}, \citenamefont {Hugonin}, \citenamefont {Kumar}, \citenamefont {Deputy}, \citenamefont {Khodkov}, \citenamefont {Rappoport}, \citenamefont {Hong}, \citenamefont {Peres}, \citenamefont {Kong}, \citenamefont {Smith},\ and\ \citenamefont {Koppens}}]{Koppens2020}%
  \BibitemOpen
  \bibfield  {author} {\bibinfo {author} {\bibfnamefont {I.}~\bibnamefont {Epstein}}, \bibinfo {author} {\bibfnamefont {D.}~\bibnamefont {Alcaraz}}, \bibinfo {author} {\bibfnamefont {Z.}~\bibnamefont {Huang}}, \bibinfo {author} {\bibfnamefont {V.-V.}\ \bibnamefont {Pusapati}}, \bibinfo {author} {\bibfnamefont {J.-P.}\ \bibnamefont {Hugonin}}, \bibinfo {author} {\bibfnamefont {A.}~\bibnamefont {Kumar}}, \bibinfo {author} {\bibfnamefont {X.~M.}\ \bibnamefont {Deputy}}, \bibinfo {author} {\bibfnamefont {T.}~\bibnamefont {Khodkov}}, \bibinfo {author} {\bibfnamefont {T.~G.}\ \bibnamefont {Rappoport}}, \bibinfo {author} {\bibfnamefont {J.-Y.}\ \bibnamefont {Hong}}, \bibinfo {author} {\bibfnamefont {N.~M.~R.}\ \bibnamefont {Peres}}, \bibinfo {author} {\bibfnamefont {J.}~\bibnamefont {Kong}}, \bibinfo {author} {\bibfnamefont {D.~R.}\ \bibnamefont {Smith}},\ and\ \bibinfo {author} {\bibfnamefont {F.~H.~L.}\ \bibnamefont {Koppens}},\ }\bibfield  {title} {\bibinfo {title} {Far-field excitation of single graphene plasmon
  cavities with ultracompressed mode volumes},\ }\href {https://doi.org/10.1126/science.abb1570} {\bibfield  {journal} {\bibinfo  {journal} {Science}\ }\textbf {\bibinfo {volume} {368}},\ \bibinfo {pages} {1219} (\bibinfo {year} {2020})}\BibitemShut {NoStop}%
\bibitem [{\citenamefont {Kaydashev}\ \emph {et~al.}(2020)\citenamefont {Kaydashev}, \citenamefont {Khlebtsov}, \citenamefont {Miakonkikh}, \citenamefont {Zhukova}, \citenamefont {Zhukov}, \citenamefont {Mylnikov}, \citenamefont {Domaratskiy},\ and\ \citenamefont {Svintsov}}]{Kaydashev2020}%
  \BibitemOpen
  \bibfield  {author} {\bibinfo {author} {\bibfnamefont {V.}~\bibnamefont {Kaydashev}}, \bibinfo {author} {\bibfnamefont {B.}~\bibnamefont {Khlebtsov}}, \bibinfo {author} {\bibfnamefont {A.}~\bibnamefont {Miakonkikh}}, \bibinfo {author} {\bibfnamefont {E.}~\bibnamefont {Zhukova}}, \bibinfo {author} {\bibfnamefont {S.}~\bibnamefont {Zhukov}}, \bibinfo {author} {\bibfnamefont {D.}~\bibnamefont {Mylnikov}}, \bibinfo {author} {\bibfnamefont {I.}~\bibnamefont {Domaratskiy}},\ and\ \bibinfo {author} {\bibfnamefont {D.}~\bibnamefont {Svintsov}},\ }\bibfield  {title} {\bibinfo {title} {Excitation of localized graphene plasmons by aperiodic self-assembled arrays of metallic antennas},\ }\href {https://doi.org/10.1088/1361-6528/aba785} {\bibfield  {journal} {\bibinfo  {journal} {Nanotechnology}\ }\textbf {\bibinfo {volume} {32}},\ \bibinfo {pages} {035201} (\bibinfo {year} {2020})}\BibitemShut {NoStop}%
\bibitem [{\citenamefont {Bandurin}\ \emph {et~al.}(2022)\citenamefont {Bandurin}, \citenamefont {M{\"o}nch}, \citenamefont {Kapralov}, \citenamefont {Phinney}, \citenamefont {Lindner}, \citenamefont {Liu}, \citenamefont {Edgar}, \citenamefont {Dmitriev}, \citenamefont {Jarillo-Herrero}, \citenamefont {Svintsov},\ and\ \citenamefont {Ganichev}}]{bandurin2022}%
  \BibitemOpen
  \bibfield  {author} {\bibinfo {author} {\bibfnamefont {D.~A.}\ \bibnamefont {Bandurin}}, \bibinfo {author} {\bibfnamefont {E.}~\bibnamefont {M{\"o}nch}}, \bibinfo {author} {\bibfnamefont {K.}~\bibnamefont {Kapralov}}, \bibinfo {author} {\bibfnamefont {I.~Y.}\ \bibnamefont {Phinney}}, \bibinfo {author} {\bibfnamefont {K.}~\bibnamefont {Lindner}}, \bibinfo {author} {\bibfnamefont {S.}~\bibnamefont {Liu}}, \bibinfo {author} {\bibfnamefont {J.~H.}\ \bibnamefont {Edgar}}, \bibinfo {author} {\bibfnamefont {I.~A.}\ \bibnamefont {Dmitriev}}, \bibinfo {author} {\bibfnamefont {P.}~\bibnamefont {Jarillo-Herrero}}, \bibinfo {author} {\bibfnamefont {D.}~\bibnamefont {Svintsov}},\ and\ \bibinfo {author} {\bibfnamefont {S.~D.}\ \bibnamefont {Ganichev}},\ }\bibfield  {title} {\bibinfo {title} {Cyclotron resonance overtones and near-field magnetoabsorption via terahertz bernstein modes in graphene},\ }\href {https://doi.org/10.1038/s41567-021-01494-8} {\bibfield  {journal} {\bibinfo  {journal} {Nat. Phys.}\ }\textbf
  {\bibinfo {volume} {18}},\ \bibinfo {pages} {462} (\bibinfo {year} {2022})}\BibitemShut {NoStop}%
\bibitem [{\citenamefont {Basov}\ \emph {et~al.}(2016)\citenamefont {Basov}, \citenamefont {Fogler},\ and\ \citenamefont {Garc{\'\i}a De~Abajo}}]{basov2016_Science}%
  \BibitemOpen
  \bibfield  {author} {\bibinfo {author} {\bibfnamefont {D.~N.}\ \bibnamefont {Basov}}, \bibinfo {author} {\bibfnamefont {M.~M.}\ \bibnamefont {Fogler}},\ and\ \bibinfo {author} {\bibfnamefont {F.~J.}\ \bibnamefont {Garc{\'\i}a De~Abajo}},\ }\bibfield  {title} {\bibinfo {title} {Polaritons in van der {W}aals materials},\ }\href {https://doi.org/10.1126/science.aag1992} {\bibfield  {journal} {\bibinfo  {journal} {Science}\ }\textbf {\bibinfo {volume} {354}},\ \bibinfo {pages} {195} (\bibinfo {year} {2016})}\BibitemShut {NoStop}%
\bibitem [{\citenamefont {Low}\ \emph {et~al.}(2017)\citenamefont {Low}, \citenamefont {Chaves}, \citenamefont {Caldwell}, \citenamefont {Kumar}, \citenamefont {Fang}, \citenamefont {Avouris}, \citenamefont {Heinz}, \citenamefont {Guinea}, \citenamefont {Martin-Moreno},\ and\ \citenamefont {Koppens}}]{low2017polaritons}%
  \BibitemOpen
  \bibfield  {author} {\bibinfo {author} {\bibfnamefont {T.}~\bibnamefont {Low}}, \bibinfo {author} {\bibfnamefont {A.}~\bibnamefont {Chaves}}, \bibinfo {author} {\bibfnamefont {J.~D.}\ \bibnamefont {Caldwell}}, \bibinfo {author} {\bibfnamefont {A.}~\bibnamefont {Kumar}}, \bibinfo {author} {\bibfnamefont {N.~X.}\ \bibnamefont {Fang}}, \bibinfo {author} {\bibfnamefont {P.}~\bibnamefont {Avouris}}, \bibinfo {author} {\bibfnamefont {T.~F.}\ \bibnamefont {Heinz}}, \bibinfo {author} {\bibfnamefont {F.}~\bibnamefont {Guinea}}, \bibinfo {author} {\bibfnamefont {L.}~\bibnamefont {Martin-Moreno}},\ and\ \bibinfo {author} {\bibfnamefont {F.}~\bibnamefont {Koppens}},\ }\bibfield  {title} {\bibinfo {title} {Polaritons in layered two-dimensional materials},\ }\href {https://doi.org/10.1038/nmat4792} {\bibfield  {journal} {\bibinfo  {journal} {Nat. Mater.}\ }\textbf {\bibinfo {volume} {16}},\ \bibinfo {pages} {182} (\bibinfo {year} {2017})}\BibitemShut {NoStop}%
\bibitem [{\citenamefont {Wang}\ \emph {et~al.}(2020)\citenamefont {Wang}, \citenamefont {Huang}, \citenamefont {Xing}, \citenamefont {Xie}, \citenamefont {Song}, \citenamefont {Wang},\ and\ \citenamefont {Yan}}]{wang2020}%
  \BibitemOpen
  \bibfield  {author} {\bibinfo {author} {\bibfnamefont {C.}~\bibnamefont {Wang}}, \bibinfo {author} {\bibfnamefont {S.}~\bibnamefont {Huang}}, \bibinfo {author} {\bibfnamefont {Q.}~\bibnamefont {Xing}}, \bibinfo {author} {\bibfnamefont {Y.}~\bibnamefont {Xie}}, \bibinfo {author} {\bibfnamefont {C.}~\bibnamefont {Song}}, \bibinfo {author} {\bibfnamefont {F.}~\bibnamefont {Wang}},\ and\ \bibinfo {author} {\bibfnamefont {H.}~\bibnamefont {Yan}},\ }\bibfield  {title} {\bibinfo {title} {Van der waals thin films of {W}{T}e$_2$ for natural hyperbolic plasmonic surfaces},\ }\href {https://doi.org/10.1038/s41467-020-15001-9} {\bibfield  {journal} {\bibinfo  {journal} {Nat. Commun.}\ }\textbf {\bibinfo {volume} {11}},\ \bibinfo {pages} {1158} (\bibinfo {year} {2020})}\BibitemShut {NoStop}%
\bibitem [{\citenamefont {de~Oliveira}\ \emph {et~al.}(2021)\citenamefont {de~Oliveira}, \citenamefont {N{\"o}renberg}, \citenamefont {{\'A}lvarez-P{\'e}rez}, \citenamefont {Wehmeier}, \citenamefont {Taboada-Guti{\'e}rrez}, \citenamefont {Obst}, \citenamefont {Hempel}, \citenamefont {Lee}, \citenamefont {Klopf}, \citenamefont {Errea}, \citenamefont {Nikitin}, \citenamefont {Kehr}, \citenamefont {Alonso-Gonz{\'a}lez},\ and\ \citenamefont {Lukas}}]{Oliveira2021}%
  \BibitemOpen
  \bibfield  {author} {\bibinfo {author} {\bibfnamefont {T.~V. A.~G.}\ \bibnamefont {de~Oliveira}}, \bibinfo {author} {\bibfnamefont {T.}~\bibnamefont {N{\"o}renberg}}, \bibinfo {author} {\bibfnamefont {G.}~\bibnamefont {{\'A}lvarez-P{\'e}rez}}, \bibinfo {author} {\bibfnamefont {L.}~\bibnamefont {Wehmeier}}, \bibinfo {author} {\bibfnamefont {J.}~\bibnamefont {Taboada-Guti{\'e}rrez}}, \bibinfo {author} {\bibfnamefont {M.}~\bibnamefont {Obst}}, \bibinfo {author} {\bibfnamefont {F.}~\bibnamefont {Hempel}}, \bibinfo {author} {\bibfnamefont {E.~J.~H.}\ \bibnamefont {Lee}}, \bibinfo {author} {\bibfnamefont {J.~M.}\ \bibnamefont {Klopf}}, \bibinfo {author} {\bibfnamefont {I.}~\bibnamefont {Errea}}, \bibinfo {author} {\bibfnamefont {A.~Y.}\ \bibnamefont {Nikitin}}, \bibinfo {author} {\bibfnamefont {S.~C.}\ \bibnamefont {Kehr}}, \bibinfo {author} {\bibfnamefont {P.}~\bibnamefont {Alonso-Gonz{\'a}lez}},\ and\ \bibinfo {author} {\bibfnamefont {M.~E.}\ \bibnamefont {Lukas}},\ }\bibfield  {title} {\bibinfo {title}
  {Nanoscale-{C}onfined {T}erahertz {P}olaritons in a van der {W}aals {C}rystal},\ }\href {https://doi.org/10.1002/adma.202005777} {\bibfield  {journal} {\bibinfo  {journal} {Adv. Mater.}\ }\textbf {\bibinfo {volume} {33}},\ \bibinfo {pages} {e2005777} (\bibinfo {year} {2021})}\BibitemShut {NoStop}%
\bibitem [{\citenamefont {Pogna}\ \emph {et~al.}(2024)\citenamefont {Pogna}, \citenamefont {Pistore}, \citenamefont {Viti}, \citenamefont {Li}, \citenamefont {Davies}, \citenamefont {Linfield},\ and\ \citenamefont {Vitiello}}]{pogna2024}%
  \BibitemOpen
  \bibfield  {author} {\bibinfo {author} {\bibfnamefont {E.~A.~A.}\ \bibnamefont {Pogna}}, \bibinfo {author} {\bibfnamefont {V.}~\bibnamefont {Pistore}}, \bibinfo {author} {\bibfnamefont {L.}~\bibnamefont {Viti}}, \bibinfo {author} {\bibfnamefont {L.}~\bibnamefont {Li}}, \bibinfo {author} {\bibfnamefont {A.~G.}\ \bibnamefont {Davies}}, \bibinfo {author} {\bibfnamefont {E.~H.}\ \bibnamefont {Linfield}},\ and\ \bibinfo {author} {\bibfnamefont {M.~S.}\ \bibnamefont {Vitiello}},\ }\bibfield  {title} {\bibinfo {title} {Near-field detection of gate-tunable anisotropic plasmon polaritons in black phosphorus at terahertz frequencies},\ }\href {https://doi.org/10.1038/s41467-024-45264-5} {\bibfield  {journal} {\bibinfo  {journal} {Nat. Commun.}\ }\textbf {\bibinfo {volume} {15}},\ \bibinfo {pages} {2373} (\bibinfo {year} {2024})}\BibitemShut {NoStop}%
\bibitem [{\citenamefont {Tsui}\ \emph {et~al.}(1980)\citenamefont {Tsui}, \citenamefont {Gornik},\ and\ \citenamefont {Logan}}]{Tsui1980}%
  \BibitemOpen
  \bibfield  {author} {\bibinfo {author} {\bibfnamefont {D.~C.}\ \bibnamefont {Tsui}}, \bibinfo {author} {\bibfnamefont {E.}~\bibnamefont {Gornik}},\ and\ \bibinfo {author} {\bibfnamefont {R.~A.}\ \bibnamefont {Logan}},\ }\bibfield  {title} {\bibinfo {title} {Far infrared emission from plasma oscillations of {S}i inversion layers},\ }\href {https://doi.org/10.1016/0038-1098(80)91043-1} {\bibfield  {journal} {\bibinfo  {journal} {Solid State Commun.}\ }\textbf {\bibinfo {volume} {35}},\ \bibinfo {pages} {875} (\bibinfo {year} {1980})}\BibitemShut {NoStop}%
\bibitem [{\citenamefont {Dyakonov}\ and\ \citenamefont {Shur}(1993)}]{Dyakonov1993}%
  \BibitemOpen
  \bibfield  {author} {\bibinfo {author} {\bibfnamefont {M.}~\bibnamefont {Dyakonov}}\ and\ \bibinfo {author} {\bibfnamefont {M.}~\bibnamefont {Shur}},\ }\bibfield  {title} {\bibinfo {title} {Shallow water analogy for a ballistic field effect transistor: New mechanism of plasma wave generation by dc current},\ }\href {https://doi.org/10.1103/PhysRevLett.71.2465} {\bibfield  {journal} {\bibinfo  {journal} {Phys. Rev. Lett.}\ }\textbf {\bibinfo {volume} {71}},\ \bibinfo {pages} {2465} (\bibinfo {year} {1993})}\BibitemShut {NoStop}%
\bibitem [{\citenamefont {Knap}\ \emph {et~al.}(2002)\citenamefont {Knap}, \citenamefont {Deng}, \citenamefont {Rumyantsev},\ and\ \citenamefont {Shur}}]{knap2002}%
  \BibitemOpen
  \bibfield  {author} {\bibinfo {author} {\bibfnamefont {W.}~\bibnamefont {Knap}}, \bibinfo {author} {\bibfnamefont {Y.}~\bibnamefont {Deng}}, \bibinfo {author} {\bibfnamefont {S.}~\bibnamefont {Rumyantsev}},\ and\ \bibinfo {author} {\bibfnamefont {M.~S.}\ \bibnamefont {Shur}},\ }\bibfield  {title} {\bibinfo {title} {Resonant detection of subterahertz and terahertz radiation by plasma waves in submicron field-effect transistors},\ }\href {https://doi.org/10.1063/1.1525851} {\bibfield  {journal} {\bibinfo  {journal} {Appl. Phys. Lett.}\ }\textbf {\bibinfo {volume} {81}},\ \bibinfo {pages} {4637} (\bibinfo {year} {2002})}\BibitemShut {NoStop}%
\bibitem [{\citenamefont {Dyakonova}\ \emph {et~al.}(2006)\citenamefont {Dyakonova}, \citenamefont {El~Fatimy}, \citenamefont {{\L}usakowski}, \citenamefont {Knap}, \citenamefont {Dyakonov}, \citenamefont {Poisson}, \citenamefont {Morvan}, \citenamefont {Bollaert}, \citenamefont {Shchepetov}, \citenamefont {Roelens}, \citenamefont {Gaquiere}, \citenamefont {Theron},\ and\ \citenamefont {Cappy}}]{Dyakonova2006}%
  \BibitemOpen
  \bibfield  {author} {\bibinfo {author} {\bibfnamefont {N.}~\bibnamefont {Dyakonova}}, \bibinfo {author} {\bibfnamefont {A.}~\bibnamefont {El~Fatimy}}, \bibinfo {author} {\bibfnamefont {J.}~\bibnamefont {{\L}usakowski}}, \bibinfo {author} {\bibfnamefont {W.}~\bibnamefont {Knap}}, \bibinfo {author} {\bibfnamefont {M.~I.}\ \bibnamefont {Dyakonov}}, \bibinfo {author} {\bibfnamefont {M.-A.}\ \bibnamefont {Poisson}}, \bibinfo {author} {\bibfnamefont {E.}~\bibnamefont {Morvan}}, \bibinfo {author} {\bibfnamefont {S.}~\bibnamefont {Bollaert}}, \bibinfo {author} {\bibfnamefont {A.}~\bibnamefont {Shchepetov}}, \bibinfo {author} {\bibfnamefont {Y.}~\bibnamefont {Roelens}}, \bibinfo {author} {\bibfnamefont {C.}~\bibnamefont {Gaquiere}}, \bibinfo {author} {\bibfnamefont {D.}~\bibnamefont {Theron}},\ and\ \bibinfo {author} {\bibfnamefont {A.}~\bibnamefont {Cappy}},\ }\bibfield  {title} {\bibinfo {title} {Room-temperature terahertz emission from nanometer field-effect transistors},\ }\href
  {https://doi.org/10.1063/1.2191421} {\bibfield  {journal} {\bibinfo  {journal} {Appl. Phys. Lett.}\ }\textbf {\bibinfo {volume} {88}},\ \bibinfo {pages} {141906} (\bibinfo {year} {2006})}\BibitemShut {NoStop}%
\bibitem [{\citenamefont {Otsuji}\ \emph {et~al.}(2008)\citenamefont {Otsuji}, \citenamefont {Meziani}, \citenamefont {Nishimura}, \citenamefont {Suemitsu}, \citenamefont {Knap}, \citenamefont {Sano}, \citenamefont {Asano},\ and\ \citenamefont {Popov}}]{otsuji2008emission}%
  \BibitemOpen
  \bibfield  {author} {\bibinfo {author} {\bibfnamefont {T.}~\bibnamefont {Otsuji}}, \bibinfo {author} {\bibfnamefont {Y.}~\bibnamefont {Meziani}}, \bibinfo {author} {\bibfnamefont {T.}~\bibnamefont {Nishimura}}, \bibinfo {author} {\bibfnamefont {T.}~\bibnamefont {Suemitsu}}, \bibinfo {author} {\bibfnamefont {W.}~\bibnamefont {Knap}}, \bibinfo {author} {\bibfnamefont {E.}~\bibnamefont {Sano}}, \bibinfo {author} {\bibfnamefont {T.}~\bibnamefont {Asano}},\ and\ \bibinfo {author} {\bibfnamefont {V.~V.}\ \bibnamefont {Popov}},\ }\bibfield  {title} {\bibinfo {title} {Emission of terahertz radiation from dual grating gate plasmon-resonant emitters fabricated with {InGaP}/{InGaAs}/{GaAs} material systems},\ }\href {https://doi.org/10.1088/0953-8984/20/38/384206} {\bibfield  {journal} {\bibinfo  {journal} {J. Phys.: Condens. Matter}\ }\textbf {\bibinfo {volume} {20}},\ \bibinfo {pages} {384206} (\bibinfo {year} {2008})}\BibitemShut {NoStop}%
\bibitem [{\citenamefont {Knap}\ \emph {et~al.}(2009)\citenamefont {Knap}, \citenamefont {Dyakonov}, \citenamefont {Coquillat}, \citenamefont {Teppe}, \citenamefont {Dyakonova}, \citenamefont {{\L}usakowski}, \citenamefont {Karpierz}, \citenamefont {Sakowicz}, \citenamefont {Valusis}, \citenamefont {Seliuta}, \citenamefont {El~Fatimy}, \citenamefont {Meziani},\ and\ \citenamefont {Otsuji}}]{knap2009}%
  \BibitemOpen
  \bibfield  {author} {\bibinfo {author} {\bibfnamefont {W.}~\bibnamefont {Knap}}, \bibinfo {author} {\bibfnamefont {M.}~\bibnamefont {Dyakonov}}, \bibinfo {author} {\bibfnamefont {D.}~\bibnamefont {Coquillat}}, \bibinfo {author} {\bibfnamefont {F.}~\bibnamefont {Teppe}}, \bibinfo {author} {\bibfnamefont {N.}~\bibnamefont {Dyakonova}}, \bibinfo {author} {\bibfnamefont {J.}~\bibnamefont {{\L}usakowski}}, \bibinfo {author} {\bibfnamefont {K.}~\bibnamefont {Karpierz}}, \bibinfo {author} {\bibfnamefont {M.}~\bibnamefont {Sakowicz}}, \bibinfo {author} {\bibfnamefont {G.}~\bibnamefont {Valusis}}, \bibinfo {author} {\bibfnamefont {D.}~\bibnamefont {Seliuta}}, \bibinfo {author} {\bibfnamefont {A.}~\bibnamefont {El~Fatimy}}, \bibinfo {author} {\bibfnamefont {Y.~M.}\ \bibnamefont {Meziani}},\ and\ \bibinfo {author} {\bibfnamefont {T.}~\bibnamefont {Otsuji}},\ }\bibfield  {title} {\bibinfo {title} {Field effect transistors for terahertz detection: Physics and first imaging applications},\ }\href
  {https://doi.org/10.1007/s10762-009-9564-9} {\bibfield  {journal} {\bibinfo  {journal} {J. Infrared Millim. Terahertz Waves}\ }\textbf {\bibinfo {volume} {30}},\ \bibinfo {pages} {1319} (\bibinfo {year} {2009})}\BibitemShut {NoStop}%
\bibitem [{\citenamefont {Muravev}\ and\ \citenamefont {Kukushkin}(2012)}]{Muravev2012}%
  \BibitemOpen
  \bibfield  {author} {\bibinfo {author} {\bibfnamefont {V.~M.}\ \bibnamefont {Muravev}}\ and\ \bibinfo {author} {\bibfnamefont {I.~V.}\ \bibnamefont {Kukushkin}},\ }\bibfield  {title} {\bibinfo {title} {Plasmonic detector/spectrometer of subterahertz radiation based on two-dimensional electron system with embedded defect},\ }\href {https://doi.org/10.1063/1.3688049} {\bibfield  {journal} {\bibinfo  {journal} {Appl. Phys. Lett.}\ }\textbf {\bibinfo {volume} {100}},\ \bibinfo {pages} {082102} (\bibinfo {year} {2012})}\BibitemShut {NoStop}%
\bibitem [{\citenamefont {Dyer}\ \emph {et~al.}(2012)\citenamefont {Dyer}, \citenamefont {Aizin}, \citenamefont {Preu}, \citenamefont {Vinh}, \citenamefont {Allen}, \citenamefont {Reno},\ and\ \citenamefont {Shaner}}]{Dyer2012}%
  \BibitemOpen
  \bibfield  {author} {\bibinfo {author} {\bibfnamefont {G.~C.}\ \bibnamefont {Dyer}}, \bibinfo {author} {\bibfnamefont {G.~R.}\ \bibnamefont {Aizin}}, \bibinfo {author} {\bibfnamefont {S.}~\bibnamefont {Preu}}, \bibinfo {author} {\bibfnamefont {N.~Q.}\ \bibnamefont {Vinh}}, \bibinfo {author} {\bibfnamefont {S.~J.}\ \bibnamefont {Allen}}, \bibinfo {author} {\bibfnamefont {J.~L.}\ \bibnamefont {Reno}},\ and\ \bibinfo {author} {\bibfnamefont {E.~A.}\ \bibnamefont {Shaner}},\ }\bibfield  {title} {\bibinfo {title} {Inducing an incipient terahertz finite plasmonic crystal in coupled two dimensional plasmonic cavities},\ }\href {https://doi.org/10.1103/PhysRevLett.109.126803} {\bibfield  {journal} {\bibinfo  {journal} {Phys. Rev. Lett.}\ }\textbf {\bibinfo {volume} {109}},\ \bibinfo {pages} {126803} (\bibinfo {year} {2012})}\BibitemShut {NoStop}%
\bibitem [{\citenamefont {Muravjov}\ \emph {et~al.}(2010)\citenamefont {Muravjov}, \citenamefont {Veksler}, \citenamefont {Popov}, \citenamefont {Polischuk}, \citenamefont {Pala}, \citenamefont {Hu}, \citenamefont {Gaska}, \citenamefont {Saxena}, \citenamefont {Peale},\ and\ \citenamefont {Shur}}]{Muravjov2010}%
  \BibitemOpen
  \bibfield  {author} {\bibinfo {author} {\bibfnamefont {A.~V.}\ \bibnamefont {Muravjov}}, \bibinfo {author} {\bibfnamefont {D.}~\bibnamefont {Veksler}}, \bibinfo {author} {\bibfnamefont {V.~V.}\ \bibnamefont {Popov}}, \bibinfo {author} {\bibfnamefont {O.~V.}\ \bibnamefont {Polischuk}}, \bibinfo {author} {\bibfnamefont {N.}~\bibnamefont {Pala}}, \bibinfo {author} {\bibfnamefont {X.}~\bibnamefont {Hu}}, \bibinfo {author} {\bibfnamefont {R.}~\bibnamefont {Gaska}}, \bibinfo {author} {\bibfnamefont {H.}~\bibnamefont {Saxena}}, \bibinfo {author} {\bibfnamefont {R.~E.}\ \bibnamefont {Peale}},\ and\ \bibinfo {author} {\bibfnamefont {M.~S.}\ \bibnamefont {Shur}},\ }\bibfield  {title} {\bibinfo {title} {Temperature dependence of plasmonic terahertz absorption in grating-gate gallium-nitride transistor structures},\ }\href {https://doi.org/10.1063/1.3292019} {\bibfield  {journal} {\bibinfo  {journal} {Appl. Phys. Lett.}\ }\textbf {\bibinfo {volume} {96}},\ \bibinfo {pages} {042105} (\bibinfo {year} {2010})}\BibitemShut
  {NoStop}%
\bibitem [{\citenamefont {Petrov}\ \emph {et~al.}(2017)\citenamefont {Petrov}, \citenamefont {Svintsov}, \citenamefont {Ryzhii},\ and\ \citenamefont {Shur}}]{Petrov2017}%
  \BibitemOpen
  \bibfield  {author} {\bibinfo {author} {\bibfnamefont {A.~S.}\ \bibnamefont {Petrov}}, \bibinfo {author} {\bibfnamefont {D.}~\bibnamefont {Svintsov}}, \bibinfo {author} {\bibfnamefont {V.}~\bibnamefont {Ryzhii}},\ and\ \bibinfo {author} {\bibfnamefont {M.~S.}\ \bibnamefont {Shur}},\ }\bibfield  {title} {\bibinfo {title} {Amplified-reflection plasmon instabilities in grating-gate plasmonic crystals},\ }\href {https://doi.org/10.1103/PhysRevB.95.045405} {\bibfield  {journal} {\bibinfo  {journal} {Phys. Rev. B}\ }\textbf {\bibinfo {volume} {95}},\ \bibinfo {pages} {045405} (\bibinfo {year} {2017})}\BibitemShut {NoStop}%
\bibitem [{\citenamefont {Kurita}\ \emph {et~al.}(2014)\citenamefont {Kurita}, \citenamefont {Ducournau}, \citenamefont {Coquillat}, \citenamefont {Satou}, \citenamefont {Kobayashi}, \citenamefont {Boubanga~Tombet}, \citenamefont {Meziani}, \citenamefont {Popov}, \citenamefont {Knap}, \citenamefont {Suemitsu},\ and\ \citenamefont {Otsuji}}]{Kurita2014}%
  \BibitemOpen
  \bibfield  {author} {\bibinfo {author} {\bibfnamefont {Y.}~\bibnamefont {Kurita}}, \bibinfo {author} {\bibfnamefont {G.}~\bibnamefont {Ducournau}}, \bibinfo {author} {\bibfnamefont {D.}~\bibnamefont {Coquillat}}, \bibinfo {author} {\bibfnamefont {A.}~\bibnamefont {Satou}}, \bibinfo {author} {\bibfnamefont {K.}~\bibnamefont {Kobayashi}}, \bibinfo {author} {\bibfnamefont {S.}~\bibnamefont {Boubanga~Tombet}}, \bibinfo {author} {\bibfnamefont {Y.~M.}\ \bibnamefont {Meziani}}, \bibinfo {author} {\bibfnamefont {V.~V.}\ \bibnamefont {Popov}}, \bibinfo {author} {\bibfnamefont {W.}~\bibnamefont {Knap}}, \bibinfo {author} {\bibfnamefont {T.}~\bibnamefont {Suemitsu}},\ and\ \bibinfo {author} {\bibfnamefont {T.}~\bibnamefont {Otsuji}},\ }\bibfield  {title} {\bibinfo {title} {Ultrahigh sensitive sub-terahertz detection by {I}n{P}-based asymmetric dual-grating-gate high-electron-mobility transistors and their broadband characteristics},\ }\href {https://doi.org/10.1063/1.4885499} {\bibfield  {journal} {\bibinfo
  {journal} {Appl. Phys. Lett.}\ }\textbf {\bibinfo {volume} {104}},\ \bibinfo {pages} {251114} (\bibinfo {year} {2014})}\BibitemShut {NoStop}%
\bibitem [{\citenamefont {Boubanga-Tombet}\ \emph {et~al.}(2014)\citenamefont {Boubanga-Tombet}, \citenamefont {Tanimoto}, \citenamefont {Satou}, \citenamefont {Suemitsu}, \citenamefont {Wang}, \citenamefont {Minamide}, \citenamefont {Ito}, \citenamefont {Fateev}, \citenamefont {Popov},\ and\ \citenamefont {Otsuji}}]{Boubanga2014current}%
  \BibitemOpen
  \bibfield  {author} {\bibinfo {author} {\bibfnamefont {S.}~\bibnamefont {Boubanga-Tombet}}, \bibinfo {author} {\bibfnamefont {Y.}~\bibnamefont {Tanimoto}}, \bibinfo {author} {\bibfnamefont {A.}~\bibnamefont {Satou}}, \bibinfo {author} {\bibfnamefont {T.}~\bibnamefont {Suemitsu}}, \bibinfo {author} {\bibfnamefont {Y.}~\bibnamefont {Wang}}, \bibinfo {author} {\bibfnamefont {H.}~\bibnamefont {Minamide}}, \bibinfo {author} {\bibfnamefont {H.}~\bibnamefont {Ito}}, \bibinfo {author} {\bibfnamefont {D.~V.}\ \bibnamefont {Fateev}}, \bibinfo {author} {\bibfnamefont {V.~V.}\ \bibnamefont {Popov}},\ and\ \bibinfo {author} {\bibfnamefont {T.}~\bibnamefont {Otsuji}},\ }\bibfield  {title} {\bibinfo {title} {Current-driven detection of terahertz radiation using a dual-grating-gate plasmonic detector},\ }\href {https://doi.org/10.1063/1.4886763} {\bibfield  {journal} {\bibinfo  {journal} {Appl. Phys. Lett.}\ }\textbf {\bibinfo {volume} {104}},\ \bibinfo {pages} {262104} (\bibinfo {year} {2014})}\BibitemShut {NoStop}%
\bibitem [{\citenamefont {Olbrich}\ \emph {et~al.}(2016)\citenamefont {Olbrich}, \citenamefont {Kamann}, \citenamefont {K\"onig}, \citenamefont {Munzert}, \citenamefont {Tutsch}, \citenamefont {Eroms}, \citenamefont {Weiss}, \citenamefont {Liu}, \citenamefont {Golub}, \citenamefont {Ivchenko}, \citenamefont {Popov}, \citenamefont {Fateev}, \citenamefont {Mashinsky}, \citenamefont {Fromm}, \citenamefont {Seyller},\ and\ \citenamefont {Ganichev}}]{Olbrich2016}%
  \BibitemOpen
  \bibfield  {author} {\bibinfo {author} {\bibfnamefont {P.}~\bibnamefont {Olbrich}}, \bibinfo {author} {\bibfnamefont {J.}~\bibnamefont {Kamann}}, \bibinfo {author} {\bibfnamefont {M.}~\bibnamefont {K\"onig}}, \bibinfo {author} {\bibfnamefont {J.}~\bibnamefont {Munzert}}, \bibinfo {author} {\bibfnamefont {L.}~\bibnamefont {Tutsch}}, \bibinfo {author} {\bibfnamefont {J.}~\bibnamefont {Eroms}}, \bibinfo {author} {\bibfnamefont {D.}~\bibnamefont {Weiss}}, \bibinfo {author} {\bibfnamefont {M.-H.}\ \bibnamefont {Liu}}, \bibinfo {author} {\bibfnamefont {L.~E.}\ \bibnamefont {Golub}}, \bibinfo {author} {\bibfnamefont {E.~L.}\ \bibnamefont {Ivchenko}}, \bibinfo {author} {\bibfnamefont {V.~V.}\ \bibnamefont {Popov}}, \bibinfo {author} {\bibfnamefont {D.~V.}\ \bibnamefont {Fateev}}, \bibinfo {author} {\bibfnamefont {K.~V.}\ \bibnamefont {Mashinsky}}, \bibinfo {author} {\bibfnamefont {F.}~\bibnamefont {Fromm}}, \bibinfo {author} {\bibfnamefont {T.}~\bibnamefont {Seyller}},\ and\ \bibinfo {author} {\bibfnamefont
  {S.~D.}\ \bibnamefont {Ganichev}},\ }\bibfield  {title} {\bibinfo {title} {Terahertz ratchet effects in graphene with a lateral superlattice},\ }\href {https://doi.org/10.1103/PhysRevB.93.075422} {\bibfield  {journal} {\bibinfo  {journal} {Phys. Rev. B}\ }\textbf {\bibinfo {volume} {93}},\ \bibinfo {pages} {075422} (\bibinfo {year} {2016})}\BibitemShut {NoStop}%
\bibitem [{\citenamefont {Boubanga-Tombet}\ \emph {et~al.}(2020)\citenamefont {Boubanga-Tombet}, \citenamefont {Knap}, \citenamefont {Yadav}, \citenamefont {Satou}, \citenamefont {But}, \citenamefont {Popov}, \citenamefont {Gorbenko}, \citenamefont {Kachorovskii},\ and\ \citenamefont {Otsuji}}]{Boubanga2020}%
  \BibitemOpen
  \bibfield  {author} {\bibinfo {author} {\bibfnamefont {S.}~\bibnamefont {Boubanga-Tombet}}, \bibinfo {author} {\bibfnamefont {W.}~\bibnamefont {Knap}}, \bibinfo {author} {\bibfnamefont {D.}~\bibnamefont {Yadav}}, \bibinfo {author} {\bibfnamefont {A.}~\bibnamefont {Satou}}, \bibinfo {author} {\bibfnamefont {D.~B.}\ \bibnamefont {But}}, \bibinfo {author} {\bibfnamefont {V.~V.}\ \bibnamefont {Popov}}, \bibinfo {author} {\bibfnamefont {I.~V.}\ \bibnamefont {Gorbenko}}, \bibinfo {author} {\bibfnamefont {V.}~\bibnamefont {Kachorovskii}},\ and\ \bibinfo {author} {\bibfnamefont {T.}~\bibnamefont {Otsuji}},\ }\bibfield  {title} {\bibinfo {title} {Room-temperature amplification of terahertz radiation by grating-gate graphene structures},\ }\href {https://doi.org/10.1103/PhysRevX.10.031004} {\bibfield  {journal} {\bibinfo  {journal} {Phys. Rev. X}\ }\textbf {\bibinfo {volume} {10}},\ \bibinfo {pages} {031004} (\bibinfo {year} {2020})}\BibitemShut {NoStop}%
\bibitem [{\citenamefont {Aizin}\ \emph {et~al.}(2023)\citenamefont {Aizin}, \citenamefont {Mikalopas},\ and\ \citenamefont {Shur}}]{Aizin2023}%
  \BibitemOpen
  \bibfield  {author} {\bibinfo {author} {\bibfnamefont {G.~R.}\ \bibnamefont {Aizin}}, \bibinfo {author} {\bibfnamefont {J.}~\bibnamefont {Mikalopas}},\ and\ \bibinfo {author} {\bibfnamefont {M.}~\bibnamefont {Shur}},\ }\bibfield  {title} {\bibinfo {title} {Plasma instability and amplified mode switching effect in {THz} field effect transistors with a grating gate},\ }\href {https://doi.org/10.1103/PhysRevB.107.245424} {\bibfield  {journal} {\bibinfo  {journal} {Phys. Rev. B}\ }\textbf {\bibinfo {volume} {107}},\ \bibinfo {pages} {245424} (\bibinfo {year} {2023})}\BibitemShut {NoStop}%
\bibitem [{\citenamefont {Sai}\ \emph {et~al.}(2023)\citenamefont {Sai}, \citenamefont {Korotyeyev}, \citenamefont {Dub}, \citenamefont {S\l{}owikowski}, \citenamefont {Filipiak}, \citenamefont {But}, \citenamefont {Ivonyak}, \citenamefont {Sakowicz}, \citenamefont {Lyaschuk}, \citenamefont {Kukhtaruk}, \citenamefont {Cywi\ifmmode~\acute{n}\else \'{n}\fi{}ski},\ and\ \citenamefont {Knap}}]{Sai2023}%
  \BibitemOpen
  \bibfield  {author} {\bibinfo {author} {\bibfnamefont {P.}~\bibnamefont {Sai}}, \bibinfo {author} {\bibfnamefont {V.~V.}\ \bibnamefont {Korotyeyev}}, \bibinfo {author} {\bibfnamefont {M.}~\bibnamefont {Dub}}, \bibinfo {author} {\bibfnamefont {M.}~\bibnamefont {S\l{}owikowski}}, \bibinfo {author} {\bibfnamefont {M.}~\bibnamefont {Filipiak}}, \bibinfo {author} {\bibfnamefont {D.~B.}\ \bibnamefont {But}}, \bibinfo {author} {\bibfnamefont {Y.}~\bibnamefont {Ivonyak}}, \bibinfo {author} {\bibfnamefont {M.}~\bibnamefont {Sakowicz}}, \bibinfo {author} {\bibfnamefont {Y.~M.}\ \bibnamefont {Lyaschuk}}, \bibinfo {author} {\bibfnamefont {S.~M.}\ \bibnamefont {Kukhtaruk}}, \bibinfo {author} {\bibfnamefont {G.}~\bibnamefont {Cywi\ifmmode~\acute{n}\else \'{n}\fi{}ski}},\ and\ \bibinfo {author} {\bibfnamefont {W.}~\bibnamefont {Knap}},\ }\bibfield  {title} {\bibinfo {title} {Electrical tuning of terahertz plasmonic crystal phases},\ }\href {https://doi.org/10.1103/PhysRevX.13.041003} {\bibfield  {journal} {\bibinfo
  {journal} {Phys. Rev. X}\ }\textbf {\bibinfo {volume} {13}},\ \bibinfo {pages} {041003} (\bibinfo {year} {2023})}\BibitemShut {NoStop}%
\bibitem [{\citenamefont {Rappoport}\ \emph {et~al.}(2020)\citenamefont {Rappoport}, \citenamefont {Epstein}, \citenamefont {Koppens},\ and\ \citenamefont {Peres}}]{rappoport2020}%
  \BibitemOpen
  \bibfield  {author} {\bibinfo {author} {\bibfnamefont {T.~G.}\ \bibnamefont {Rappoport}}, \bibinfo {author} {\bibfnamefont {I.}~\bibnamefont {Epstein}}, \bibinfo {author} {\bibfnamefont {F.~H.~L.}\ \bibnamefont {Koppens}},\ and\ \bibinfo {author} {\bibfnamefont {N.~M.~R.}\ \bibnamefont {Peres}},\ }\bibfield  {title} {\bibinfo {title} {Understanding the electromagnetic response of graphene/metallic nanostructures hybrids of different dimensionality},\ }\href {https://doi.org/10.1021/acsphotonics.0c01002} {\bibfield  {journal} {\bibinfo  {journal} {ACS Photonics}\ }\textbf {\bibinfo {volume} {7}},\ \bibinfo {pages} {2302} (\bibinfo {year} {2020})}\BibitemShut {NoStop}%
\bibitem [{\citenamefont {Zabolotnykh}\ and\ \citenamefont {Volkov}(2019{\natexlab{a}})}]{zabolotnykh2019}%
  \BibitemOpen
  \bibfield  {author} {\bibinfo {author} {\bibfnamefont {A.~A.}\ \bibnamefont {Zabolotnykh}}\ and\ \bibinfo {author} {\bibfnamefont {V.~A.}\ \bibnamefont {Volkov}},\ }\bibfield  {title} {\bibinfo {title} {Interaction of gated and ungated plasmons in two-dimensional electron systems},\ }\href {https://doi.org/10.1103/PhysRevB.99.165304} {\bibfield  {journal} {\bibinfo  {journal} {Phys. Rev. B}\ }\textbf {\bibinfo {volume} {99}},\ \bibinfo {pages} {165304} (\bibinfo {year} {2019}{\natexlab{a}})}\BibitemShut {NoStop}%
\bibitem [{\citenamefont {Muravev}\ \emph {et~al.}(2019{\natexlab{a}})\citenamefont {Muravev}, \citenamefont {Gusikhin}, \citenamefont {Zarezin}, \citenamefont {Andreev}, \citenamefont {Gubarev},\ and\ \citenamefont {Kukushkin}}]{Muravev2019_2Dstripe}%
  \BibitemOpen
  \bibfield  {author} {\bibinfo {author} {\bibfnamefont {V.~M.}\ \bibnamefont {Muravev}}, \bibinfo {author} {\bibfnamefont {P.~A.}\ \bibnamefont {Gusikhin}}, \bibinfo {author} {\bibfnamefont {A.~M.}\ \bibnamefont {Zarezin}}, \bibinfo {author} {\bibfnamefont {I.~V.}\ \bibnamefont {Andreev}}, \bibinfo {author} {\bibfnamefont {S.~I.}\ \bibnamefont {Gubarev}},\ and\ \bibinfo {author} {\bibfnamefont {I.~V.}\ \bibnamefont {Kukushkin}},\ }\bibfield  {title} {\bibinfo {title} {Two-dimensional plasmon induced by metal proximity},\ }\href {https://doi.org/10.1103/PhysRevB.99.241406} {\bibfield  {journal} {\bibinfo  {journal} {Phys. Rev. B}\ }\textbf {\bibinfo {volume} {99}},\ \bibinfo {pages} {241406(R)} (\bibinfo {year} {2019}{\natexlab{a}})}\BibitemShut {NoStop}%
\bibitem [{\citenamefont {Zarezin}\ \emph {et~al.}(2020)\citenamefont {Zarezin}, \citenamefont {Gusikhin}, \citenamefont {Muravev},\ and\ \citenamefont {Kukushkin}}]{zarezin2020}%
  \BibitemOpen
  \bibfield  {author} {\bibinfo {author} {\bibfnamefont {A.~M.}\ \bibnamefont {Zarezin}}, \bibinfo {author} {\bibfnamefont {P.~A.}\ \bibnamefont {Gusikhin}}, \bibinfo {author} {\bibfnamefont {V.~M.}\ \bibnamefont {Muravev}},\ and\ \bibinfo {author} {\bibfnamefont {I.~V.}\ \bibnamefont {Kukushkin}},\ }\bibfield  {title} {\bibinfo {title} {Spectra of two-dimensional “proximity” plasmons measured by the standing-wave method},\ }\href {https://doi.org/10.1134/S0021364020050112} {\bibfield  {journal} {\bibinfo  {journal} {JETP Lett.}\ }\textbf {\bibinfo {volume} {111}},\ \bibinfo {pages} {282} (\bibinfo {year} {2020})}\BibitemShut {NoStop}%
\bibitem [{\citenamefont {Zabolotnykh}\ and\ \citenamefont {Volkov}(2019{\natexlab{b}})}]{zabolotnykh2019plasmons}%
  \BibitemOpen
  \bibfield  {author} {\bibinfo {author} {\bibfnamefont {A.~A.}\ \bibnamefont {Zabolotnykh}}\ and\ \bibinfo {author} {\bibfnamefont {V.~A.}\ \bibnamefont {Volkov}},\ }\bibfield  {title} {\bibinfo {title} {Plasmons in infinite 2{D} electron system screened by the disk-shaped metallic gate},\ }\href {https://doi.org/10.1134/S1063782619140276} {\bibfield  {journal} {\bibinfo  {journal} {Semiconductors}\ }\textbf {\bibinfo {volume} {53}},\ \bibinfo {pages} {1870} (\bibinfo {year} {2019}{\natexlab{b}})}\BibitemShut {NoStop}%
\bibitem [{\citenamefont {Zabolotnykh}\ \emph {et~al.}(2021)\citenamefont {Zabolotnykh}, \citenamefont {Enaldiev},\ and\ \citenamefont {Volkov}}]{zabolotnykh2021}%
  \BibitemOpen
  \bibfield  {author} {\bibinfo {author} {\bibfnamefont {A.~A.}\ \bibnamefont {Zabolotnykh}}, \bibinfo {author} {\bibfnamefont {V.~V.}\ \bibnamefont {Enaldiev}},\ and\ \bibinfo {author} {\bibfnamefont {V.~A.}\ \bibnamefont {Volkov}},\ }\bibfield  {title} {\bibinfo {title} {Quasistationary near-gate plasmons in van der {W}aals heterostructures},\ }\href {https://doi.org/10.1103/PhysRevB.104.195435} {\bibfield  {journal} {\bibinfo  {journal} {Phys. Rev. B}\ }\textbf {\bibinfo {volume} {104}},\ \bibinfo {pages} {195435} (\bibinfo {year} {2021})}\BibitemShut {NoStop}%
\bibitem [{\citenamefont {Muravev}\ \emph {et~al.}(2019{\natexlab{b}})\citenamefont {Muravev}, \citenamefont {Zarezin}, \citenamefont {Gusikhin}, \citenamefont {Shupletsov},\ and\ \citenamefont {Kukushkin}}]{Muravev_ZarezinPRB2019}%
  \BibitemOpen
  \bibfield  {author} {\bibinfo {author} {\bibfnamefont {V.~M.}\ \bibnamefont {Muravev}}, \bibinfo {author} {\bibfnamefont {A.~M.}\ \bibnamefont {Zarezin}}, \bibinfo {author} {\bibfnamefont {P.~A.}\ \bibnamefont {Gusikhin}}, \bibinfo {author} {\bibfnamefont {A.~V.}\ \bibnamefont {Shupletsov}},\ and\ \bibinfo {author} {\bibfnamefont {I.~V.}\ \bibnamefont {Kukushkin}},\ }\bibfield  {title} {\bibinfo {title} {Proximity plasma excitations in disk and ring geometries},\ }\href {https://doi.org/10.1103/PhysRevB.100.205405} {\bibfield  {journal} {\bibinfo  {journal} {Phys. Rev. B}\ }\textbf {\bibinfo {volume} {100}},\ \bibinfo {pages} {205405} (\bibinfo {year} {2019}{\natexlab{b}})}\BibitemShut {NoStop}%
\bibitem [{\citenamefont {Muravev}\ \emph {et~al.}(2021)\citenamefont {Muravev}, \citenamefont {Andreev}, \citenamefont {Semenov}, \citenamefont {Gubarev},\ and\ \citenamefont {Kukushkin}}]{muravev2021crossover}%
  \BibitemOpen
  \bibfield  {author} {\bibinfo {author} {\bibfnamefont {V.~M.}\ \bibnamefont {Muravev}}, \bibinfo {author} {\bibfnamefont {I.~V.}\ \bibnamefont {Andreev}}, \bibinfo {author} {\bibfnamefont {N.~D.}\ \bibnamefont {Semenov}}, \bibinfo {author} {\bibfnamefont {S.~I.}\ \bibnamefont {Gubarev}},\ and\ \bibinfo {author} {\bibfnamefont {I.~V.}\ \bibnamefont {Kukushkin}},\ }\bibfield  {title} {\bibinfo {title} {Crossover from proximity to ordinary two-dimensional plasma excitation},\ }\href {https://doi.org/10.1103/PhysRevB.103.125308} {\bibfield  {journal} {\bibinfo  {journal} {Phys. Rev. B}\ }\textbf {\bibinfo {volume} {103}},\ \bibinfo {pages} {125308} (\bibinfo {year} {2021})}\BibitemShut {NoStop}%
\bibitem [{\citenamefont {Zarezin}\ \emph {et~al.}(2021)\citenamefont {Zarezin}, \citenamefont {Gusikhin}, \citenamefont {Andreev}, \citenamefont {Muravev},\ and\ \citenamefont {Kukushkin}}]{zarezin2021review}%
  \BibitemOpen
  \bibfield  {author} {\bibinfo {author} {\bibfnamefont {A.~M.}\ \bibnamefont {Zarezin}}, \bibinfo {author} {\bibfnamefont {P.~A.}\ \bibnamefont {Gusikhin}}, \bibinfo {author} {\bibfnamefont {I.~V.}\ \bibnamefont {Andreev}}, \bibinfo {author} {\bibfnamefont {V.~M.}\ \bibnamefont {Muravev}},\ and\ \bibinfo {author} {\bibfnamefont {I.~V.}\ \bibnamefont {Kukushkin}},\ }\bibfield  {title} {\bibinfo {title} {Plasmon excitations in partially screened two-dimensional electron systems (brief review)},\ }\href {https://doi.org/10.1134/S0021364021110096} {\bibfield  {journal} {\bibinfo  {journal} {JETP Lett.}\ }\textbf {\bibinfo {volume} {113}},\ \bibinfo {pages} {713} (\bibinfo {year} {2021})}\BibitemShut {NoStop}%
\bibitem [{\citenamefont {Dyakonov}(2008)}]{Dyakonov2008}%
  \BibitemOpen
  \bibfield  {author} {\bibinfo {author} {\bibfnamefont {M.~I.}\ \bibnamefont {Dyakonov}},\ }\bibfield  {title} {\bibinfo {title} {Boundary instability of a two-dimensional electron fluid},\ }\href {https://doi.org/10.1134/S1063782608080186} {\bibfield  {journal} {\bibinfo  {journal} {Semiconductors}\ }\textbf {\bibinfo {volume} {42}},\ \bibinfo {pages} {984} (\bibinfo {year} {2008})}\BibitemShut {NoStop}%
\bibitem [{\citenamefont {Petrov}\ \emph {et~al.}(2016)\citenamefont {Petrov}, \citenamefont {Svintsov}, \citenamefont {Rudenko}, \citenamefont {Ryzhii},\ and\ \citenamefont {Shur}}]{Petrov2016}%
  \BibitemOpen
  \bibfield  {author} {\bibinfo {author} {\bibfnamefont {A.~S.}\ \bibnamefont {Petrov}}, \bibinfo {author} {\bibfnamefont {D.}~\bibnamefont {Svintsov}}, \bibinfo {author} {\bibfnamefont {M.}~\bibnamefont {Rudenko}}, \bibinfo {author} {\bibfnamefont {V.}~\bibnamefont {Ryzhii}},\ and\ \bibinfo {author} {\bibfnamefont {M.~S.}\ \bibnamefont {Shur}},\ }\bibfield  {title} {\bibinfo {title} {Plasma instability of 2{D} electrons in a field effect transistor with a partly gated channel},\ }\href {https://doi.org/10.1142/S0129156416400152} {\bibfield  {journal} {\bibinfo  {journal} {Int. J. High Speed Electron. Syst.}\ }\textbf {\bibinfo {volume} {25}},\ \bibinfo {pages} {1640015} (\bibinfo {year} {2016})}\BibitemShut {NoStop}%
\bibitem [{\citenamefont {Satou}\ \emph {et~al.}(2003)\citenamefont {Satou}, \citenamefont {Khmyrova}, \citenamefont {Ryzhii},\ and\ \citenamefont {Shur}}]{Satou2003}%
  \BibitemOpen
  \bibfield  {author} {\bibinfo {author} {\bibfnamefont {A.}~\bibnamefont {Satou}}, \bibinfo {author} {\bibfnamefont {I.}~\bibnamefont {Khmyrova}}, \bibinfo {author} {\bibfnamefont {V.}~\bibnamefont {Ryzhii}},\ and\ \bibinfo {author} {\bibfnamefont {M.~S.}\ \bibnamefont {Shur}},\ }\bibfield  {title} {\bibinfo {title} {Plasma and transit-time mechanisms of the terahertz radiation detection in high-electron-mobility transistors},\ }\href {https://doi.org/10.1088/0268-1242/18/6/312} {\bibfield  {journal} {\bibinfo  {journal} {Semicond. Sci. Technol.}\ }\textbf {\bibinfo {volume} {18}},\ \bibinfo {pages} {460} (\bibinfo {year} {2003})}\BibitemShut {NoStop}%
\bibitem [{\citenamefont {Dyakonov}\ and\ \citenamefont {Shur}(2005)}]{Dyakonov2005}%
  \BibitemOpen
  \bibfield  {author} {\bibinfo {author} {\bibfnamefont {M.}~\bibnamefont {Dyakonov}}\ and\ \bibinfo {author} {\bibfnamefont {M.~S.}\ \bibnamefont {Shur}},\ }\bibfield  {title} {\bibinfo {title} {Current instability and plasma waves generation in ungated two-dimensional electron layers},\ }\href {https://doi.org/10.1063/1.2042547} {\bibfield  {journal} {\bibinfo  {journal} {Appl. Phys. Lett.}\ }\textbf {\bibinfo {volume} {87}},\ \bibinfo {pages} {111501} (\bibinfo {year} {2005})}\BibitemShut {NoStop}%
\bibitem [{\citenamefont {Jiang}\ \emph {et~al.}(2018)\citenamefont {Jiang}, \citenamefont {Mele},\ and\ \citenamefont {Fogler}}]{Jiang2018}%
  \BibitemOpen
  \bibfield  {author} {\bibinfo {author} {\bibfnamefont {B.-Y.}\ \bibnamefont {Jiang}}, \bibinfo {author} {\bibfnamefont {E.~J.}\ \bibnamefont {Mele}},\ and\ \bibinfo {author} {\bibfnamefont {M.~M.}\ \bibnamefont {Fogler}},\ }\bibfield  {title} {\bibinfo {title} {Theory of plasmon reflection by a 1{D} junction},\ }\href {https://doi.org/10.1364/OE.26.017209} {\bibfield  {journal} {\bibinfo  {journal} {Opt. Express}\ }\textbf {\bibinfo {volume} {26}},\ \bibinfo {pages} {17209} (\bibinfo {year} {2018})}\BibitemShut {NoStop}%
\bibitem [{\citenamefont {Rejaei}\ and\ \citenamefont {Khavasi}(2015)}]{rejaei2015}%
  \BibitemOpen
  \bibfield  {author} {\bibinfo {author} {\bibfnamefont {B.}~\bibnamefont {Rejaei}}\ and\ \bibinfo {author} {\bibfnamefont {A.}~\bibnamefont {Khavasi}},\ }\bibfield  {title} {\bibinfo {title} {Scattering of surface plasmons on graphene by a discontinuity in surface conductivity},\ }\href {https://doi.org/10.1088/2040-8978/17/7/075002} {\bibfield  {journal} {\bibinfo  {journal} {J. Opt.}\ }\textbf {\bibinfo {volume} {17}},\ \bibinfo {pages} {075002} (\bibinfo {year} {2015})}\BibitemShut {NoStop}%
\bibitem [{\citenamefont {Siaber}\ \emph {et~al.}(2019)\citenamefont {Siaber}, \citenamefont {Zonetti},\ and\ \citenamefont {Sydoruk}}]{siaber2019}%
  \BibitemOpen
  \bibfield  {author} {\bibinfo {author} {\bibfnamefont {S.}~\bibnamefont {Siaber}}, \bibinfo {author} {\bibfnamefont {S.}~\bibnamefont {Zonetti}},\ and\ \bibinfo {author} {\bibfnamefont {O.}~\bibnamefont {Sydoruk}},\ }\bibfield  {title} {\bibinfo {title} {Junctions between two-dimensional plasmonic waveguides in the presence of retardation},\ }\href {https://doi.org/10.1088/2040-8986/ab4056} {\bibfield  {journal} {\bibinfo  {journal} {J. Opt.}\ }\textbf {\bibinfo {volume} {21}},\ \bibinfo {pages} {105002} (\bibinfo {year} {2019})}\BibitemShut {NoStop}%
\bibitem [{\citenamefont {Aizin}\ and\ \citenamefont {Dyer}(2012)}]{Aizin2012}%
  \BibitemOpen
  \bibfield  {author} {\bibinfo {author} {\bibfnamefont {G.~R.}\ \bibnamefont {Aizin}}\ and\ \bibinfo {author} {\bibfnamefont {G.~C.}\ \bibnamefont {Dyer}},\ }\bibfield  {title} {\bibinfo {title} {Transmission line theory of collective plasma excitations in periodic two-dimensional electron systems: Finite plasmonic crystals and {T}amm states},\ }\href {https://doi.org/10.1103/PhysRevB.86.235316} {\bibfield  {journal} {\bibinfo  {journal} {Phys. Rev. B}\ }\textbf {\bibinfo {volume} {86}},\ \bibinfo {pages} {235316} (\bibinfo {year} {2012})}\BibitemShut {NoStop}%
\bibitem [{Note1()}]{Note1}%
  \BibitemOpen
  \bibinfo {note} {Indeed, plasmons in a 2DES with a single strip-shaped gate were studied in Ref.~\cite {zabolotnykh2019} reporting that the electron charge density is almost entirely localized in the gated area of the 2DES, being absent in the ungated regions, see Figs.~4(a) and (b) in Ref.~\cite {zabolotnykh2019}. For the 2DES with a grating gate considered in this paper, the same behavior of the 2D charge density is expected as the distance $d$ between the 2DES and the gate is small compared to other characteristic lengths, such as $b$, $W$, $q_y^{-1}$, etc. Thus, electrons in the gated areas of the 2DES interact strongly with their 'images' induced in the corresponding metal gates, whereas the interactions of 2D electrons under different gates or electrons in different gates are weak. Therefore, in the limit of small $d$, the description of electron dynamics in the 2DES with a grating gate is very similar to the case of a 2DES with a single gate. It should also be noted that in the framework of the
  employed (optical) approach, the interaction between the electrons in different gates and 2D electrons under different gates is neglected.}\BibitemShut {Stop}%
\bibitem [{Note2()}]{Note2}%
  \BibitemOpen
  \bibinfo {note} {For details, see the Discussion section in Ref.~\cite {zabolotnykh2021}.}\BibitemShut {Stop}%
\bibitem [{\citenamefont {Mast}\ \emph {et~al.}(1985)\citenamefont {Mast}, \citenamefont {Dahm},\ and\ \citenamefont {Fetter}}]{Mast1985}%
  \BibitemOpen
  \bibfield  {author} {\bibinfo {author} {\bibfnamefont {D.~B.}\ \bibnamefont {Mast}}, \bibinfo {author} {\bibfnamefont {A.~J.}\ \bibnamefont {Dahm}},\ and\ \bibinfo {author} {\bibfnamefont {A.~L.}\ \bibnamefont {Fetter}},\ }\bibfield  {title} {\bibinfo {title} {Observation of bulk and edge magnetoplasmons in a two-dimensional electron fluid},\ }\href {https://doi.org/10.1103/PhysRevLett.54.1706} {\bibfield  {journal} {\bibinfo  {journal} {Phys. Rev. Lett.}\ }\textbf {\bibinfo {volume} {54}},\ \bibinfo {pages} {1706} (\bibinfo {year} {1985})}\BibitemShut {NoStop}%
\bibitem [{\citenamefont {Glattli}\ \emph {et~al.}(1985)\citenamefont {Glattli}, \citenamefont {Andrei}, \citenamefont {Deville}, \citenamefont {Poitrenaud},\ and\ \citenamefont {Williams}}]{Glattli1985}%
  \BibitemOpen
  \bibfield  {author} {\bibinfo {author} {\bibfnamefont {D.~C.}\ \bibnamefont {Glattli}}, \bibinfo {author} {\bibfnamefont {E.~Y.}\ \bibnamefont {Andrei}}, \bibinfo {author} {\bibfnamefont {G.}~\bibnamefont {Deville}}, \bibinfo {author} {\bibfnamefont {J.}~\bibnamefont {Poitrenaud}},\ and\ \bibinfo {author} {\bibfnamefont {F.~I.~B.}\ \bibnamefont {Williams}},\ }\bibfield  {title} {\bibinfo {title} {Dynamical hall effect in a two-dimensional classical plasma},\ }\href {https://doi.org/10.1103/PhysRevLett.54.1710} {\bibfield  {journal} {\bibinfo  {journal} {Phys. Rev. Lett.}\ }\textbf {\bibinfo {volume} {54}},\ \bibinfo {pages} {1710} (\bibinfo {year} {1985})}\BibitemShut {NoStop}%
\bibitem [{\citenamefont {Volkov}\ and\ \citenamefont {Mikhailov}(1985)}]{Volkov1985}%
  \BibitemOpen
  \bibfield  {author} {\bibinfo {author} {\bibfnamefont {V.~A.}\ \bibnamefont {Volkov}}\ and\ \bibinfo {author} {\bibfnamefont {S.~A.}\ \bibnamefont {Mikhailov}},\ }\bibfield  {title} {\bibinfo {title} {Theory of edge magnetoplasmons in a two-dimensional electron gas},\ }\href {http://jetpletters.ru/ps/1439/article_21888.pdf} {\bibfield  {journal} {\bibinfo  {journal} {Pis’ma Zh. Eksp. Teor. Fiz.}\ }\textbf {\bibinfo {volume} {42}},\ \bibinfo {pages} {450} (\bibinfo {year} {1985})},\ \translation{JETP Lett. \textbf{42}, 556 (1985)}\BibitemShut {NoStop}%
\bibitem [{\citenamefont {Volkov}\ and\ \citenamefont {Mikhailov}(1988)}]{Volkov1988}%
  \BibitemOpen
  \bibfield  {author} {\bibinfo {author} {\bibfnamefont {V.~A.}\ \bibnamefont {Volkov}}\ and\ \bibinfo {author} {\bibfnamefont {S.~A.}\ \bibnamefont {Mikhailov}},\ }\bibfield  {title} {\bibinfo {title} {Edge magnetoplasmons: low frequency weakly damped excitations in inhomogeneous two-dimensional electron systems},\ }\href {http://www.jetp.ras.ru/cgi-bin/dn/e_067_08_1639.pdf} {\bibfield  {journal} {\bibinfo  {journal} {Zh. Eksp. Teor. Fiz.}\ }\textbf {\bibinfo {volume} {94}},\ \bibinfo {pages} {217} (\bibinfo {year} {1988})},\ \translation{Sov. Phys. JETP \textbf{67}, 1639 (1988)}\BibitemShut {NoStop}%
\bibitem [{\citenamefont {Heiblum}\ \emph {et~al.}(1984)\citenamefont {Heiblum}, \citenamefont {Mendez},\ and\ \citenamefont {Stern}}]{heiblum1984}%
  \BibitemOpen
  \bibfield  {author} {\bibinfo {author} {\bibfnamefont {M.}~\bibnamefont {Heiblum}}, \bibinfo {author} {\bibfnamefont {E.~E.}\ \bibnamefont {Mendez}},\ and\ \bibinfo {author} {\bibfnamefont {F.}~\bibnamefont {Stern}},\ }\bibfield  {title} {\bibinfo {title} {High mobility electron gas in selectively doped n: {A}l{G}a{A}s/{G}a{A}s heterojunctions},\ }\href {https://doi.org/10.1063/1.94644} {\bibfield  {journal} {\bibinfo  {journal} {Appl. Phys. Lett.}\ }\textbf {\bibinfo {volume} {44}},\ \bibinfo {pages} {1064} (\bibinfo {year} {1984})}\BibitemShut {NoStop}%
\end{thebibliography}%

\end{document}